\documentclass[12pt]{article}

\usepackage{epsfig,latexsym,amsfonts,amsmath,amsthm,amssymb,amsbsy,multirow,slashed,wasysym,textcomp,comment,mathtools,cancel,bbold,cite,datetime}
\usepackage[export]{adjustbox}
\usepackage{geometry}
 \usepackage{pgfplots}
 \usepgfplotslibrary{patchplots}
 \pgfplotsset{compat=newest}
\usepackage{mathrsfs}

\theoremstyle{definition}

\newtheorem{theorem}{Theorem}
\DeclareGraphicsExtensions{.pdf,.png}
\usetikzlibrary{patterns}
\usepackage{tikz}
\usetikzlibrary{arrows.meta}
\usetikzlibrary{shapes.symbols}
\usetikzlibrary{calc}
\tikzstyle{bag} = [align=center]
\usetikzlibrary{decorations.pathmorphing}
\usepackage[normalem]{ulem}
\usepackage[shortlabels]{enumitem}

\usepackage{tocloft}
\usepackage{bbold}
\usepackage{hyperref}
 \hypersetup{
      linktoc=all,    
  }
\usepackage[version=3]{mhchem}

\usepackage{pgfplots}

 \newcommand{\badat}{\begin{alignedat}}
 \newcommand{\eadat}{\end{alignedat}}
 \newcommand\scalemath[2]{\scalebox{#1}{\mbox{\ensuremath{\displaystyle #2}}}}
 \def\be{\begin{equation}}
\def\ee{\end{equation}}

\def\p{\partial}

\usepackage{color}

\newcommand{\pink}[1]{\textcolor{\pink}{#1}}

\definecolor{dblue}{rgb}{0.2,0.50,0.80}

\def\C{\mathcal{C}}

\def\A{\mathcal{A}}

\def\U{\mathcal{U}}

\def\bz{{\bar z}}

\def\ba{{\bar a}}
\def\bb{{\bar b}}
\def\bc{{\bar c}}
\def\bd{{\bar d}}

\def\bz{{\bar z}}

\def\R{\mathbb{R}}
\def\C{\mathbb{C}}
\def\CP{\mathbb{CP}}
\def\Z{\mathbb{Z}}
\def\SU{\mathrm{SU}}
\def\U{\mathrm{1}}
\def\SL{\mathrm{SL}}
\def\ISO{\mathrm{ISO}}
\def\d{\mathrm{d}}
\def\Im{\mathrm{Im}}

\def\U{\mathcal{U}}

\usepackage{tkz-euclide}
\usetikzlibrary{calc,through}

\numberwithin{equation}{section} 

\begin{document}

 \begin{titlepage}
  \thispagestyle{empty}
  \begin{flushright}
  \end{flushright}
  \bigskip
  \begin{center}

        \baselineskip=13pt {\huge {
     Celestial Geometry}}
     
      \vskip1cm 

   \centerline{ {Sebastian Mizera}${}^{1}$ and
   {Sabrina Pasterski}${}^{2,3}$
}
\def \etatild {\tilde{\eta}}

\bigskip\bigskip
 
 \centerline{\em${}^1$ Institute for Advanced Study, Princeton, NJ 08540, USA}

 \bigskip

 \centerline{\em${}^2$ Princeton Center for Theoretical Science, Princeton, NJ 08544, USA}

 \bigskip

  \centerline{\em${}^3$   Perimeter Institute for Theoretical Physics, Waterloo, ON N2L 2Y5, Canada
 }

\bigskip\bigskip

\end{center}

\begin{abstract}
 \noindent

Celestial holography expresses $\mathcal{S}$-matrix elements as correlators in a CFT living on the night sky. Poincar\'e invariance imposes additional selection rules on the allowed positions of operators. As a consequence, $n$-point correlators are only supported on certain patches of the celestial sphere, depending on the labeling of each operator as incoming/outgoing. Here we initiate a study of the {\it celestial geometry}, examining the kinematic support of celestial amplitudes for different crossing channels. We give simple geometric rules for determining this support. For $n\ge 5$, we can view these channels as tiling together to form a covering of the celestial sphere. Our analysis serves as a stepping off point to better understand the analyticity of celestial correlators and illuminate the connection between the 4D kinematic and 2D CFT notions of crossing symmetry.

\end{abstract}

\end{titlepage}

\tableofcontents

\section{Introduction}

Recent efforts to understand the holographic nature of quantum gravity for vanishing cosmological constant have led to an exciting merger of techniques from the relativity, conformal bootstrap, and amplitudes communities~\cite{Pasterski:2021raf}. If we attempt to replicate existing holographic dictionaries~\cite{Maldacena:1997re,Witten:1998qj} from the ground up by matching symmetries~\cite{Brown:1986nw,Brown:1992br}, we are naturally led to broaden our scope from the Poincar\'e isometries of Minkowski space to the asymptotic symmetries of asymptotically flat spacetimes~\cite{Bondi:1962px,Sachs:1962wk,Sachs:1962zza}.  These include extensions of the Lorentz group to a Virasoro symmetry~\cite{Barnich:2009se,Barnich:2011ct,Cachazo:2014fwa,Kapec:2014opa,Kapec:2016jld} and hint at a CFT living on the night sky.  We can realize this by recasting the $\mathcal{S}$-matrix program in terms of a dual `celestial CFT' (CCFT), wherein 4D $\mathcal{S}$-matrix elements are mapped to 2D correlators of primary operators.  The act of bootstrapping correlators based on the symmetries and OPE data amounts to bootstrapping amplitudes from their soft and collinear limits. Discovering an intrinsic description of this dual would be tantamount to determining rules for the on-shell data of `consistent' bulk theories~\cite{Eden:1966dnq,Arkani-Hamed:2006emk,Arkani-Hamed:2020blm,Arkani-Hamed:2021ajd}.

The holographic map is implemented by a change of basis from plane-wave scattering to boost eigenstates. Let us briefly set up our conventions for what follows.  Throughout this paper we will consider massless scattering in four spacetime dimensions. The $n$ external momenta take the form
\be\label{pi}
p_i^\mu = \epsilon_i \omega_i \big( 1 + z_i \bar{z}_i, \; z_i + \bar{z}_i, \; -i(z_i - \bar{z}_i),\; 1 - z_i \bar{z}_i \big)
\ee
where the sign $ \epsilon_i=\pm1 $ determines if the particle is incoming or outgoing. Recall that the $n$-point scattering amplitude $\mathcal{A}_n(p_i)$ is a distribution containing the momentum-conservation delta function $\delta^{4}({\textstyle\sum}_{i=1}^{n} p_i)$. If we now view the Lorentz group in 4D as conformal transformations of the celestial sphere,  the helicity of an external scattering state determines the 2D spin $J_i$. We can further trade the energy variable $\omega_i$ for a conformal dimension $\Delta_i$ by performing a Mellin transform\footnote{
Throughout this paper we will use the term `celestial amplitude' to refer to the Mellin transformed amplitude~\eqref{Acele}. There are generalizations~\cite{Pasterski:2017kqt,ss,Atanasov:2021cje,Sharma:2021gcz} which involve integral transforms on the celestial sphere that act as intertwiners, taking us between Weyl reflected SL($2,\mathbb{C})$ representations. 
The lessons we learn about amplitudes in the Mellin basis can be translated to their smeared analogs by applying the respective additional transformations. 
Some historical context for this Lorentz basis and other options that have yet to be explored in the modern celestial literature can be found in appendix~\ref{app:LorentzBasis}.}
\be\label{Acele}
	\tilde{\A}_n (\Delta_i,z_i,\bz_i)=
	\left(\prod_{i=1}^{n} \mathcal{N}_{\Delta_i,\epsilon_i} \int_0^\infty \d\omega\, \omega^{\Delta_i-1} \right)\mathcal{A}_n (p_i).
\ee
In what follows we will take $ \mathcal{N}_{\Delta_i,\epsilon_i} =1$ since we will see later that this choice has nice crossing properties. While this map guarantees that the external scattering states transform as quasi-primaries in a 2D CFT, these get promoted to Virasoro primaries upon coupling to gravity~\cite{Kapec:2014opa,Kapec:2016jld,Ball:2019atb,Pasterski:2022lsl}. The resultant `celestial CFT' seems to posses rich and intriguing properties unfamiliar from regular two-dimensional CFTs. For example, its spectrum involves states with complex conformal dimensions, with finite energy scattering is captured by conformal dimensions on the principal series $\Delta_i=1+i\lambda_i$~\cite{deBoer:2003vf,Pasterski:2017kqt}.

\begin{figure}[t]
\begin{center}
\begin{tikzpicture}[scale=1.25, every node/.style={scale=.9}]
\draw[gray] (0,0) -- (7,0) -- (8.5,2) -- (1.5,2) -- (0,0);
\draw[dashed, thick, rotate=3] (4.2,.8) ellipse (5.1em and 2em) ;
\draw[ultra thick] (4.15,.9) circle (5.1 em);
\node[] at (6.5,3) {$\mathcal{CS}^2$};
\draw[fill] (4.15-2.1,.9) circle (.2em) node[left,red] {$z_1~$};
\draw[fill] (4.15+2.1,.9) circle (.2em) node[right,red] {$~z_2$};
\draw[gray, dotted]  (4.15-2.1,.9)--(4.15+2.1,.9);
\draw[fill] (3.3,.2) circle (.2em) node[below,cyan!70!blue] {\raisebox{-4mm}{$z_4$}};
\draw[fill] (3.3+2.1,1.75) circle (.2em) node[above,cyan!70!blue] {\raisebox{3mm}{$z_3$}};
\draw[gray, dotted]  (3.3,.2)--(3.3+2.1,1.75);
\coordinate (a) at (4.25,.9);
\coordinate (b) at (3.3+2.1,1.75);
\coordinate (c) at (3.3,.2);
\coordinate (d) at (4.15-2.1,.9);
\coordinate (e) at (4.15+2.1,.9);
\draw[thick, -Straight Barb] ($(a)!0.10!(b)$)-- node[above left,cyan!70!blue] {\raisebox{-2mm}{$p_3$}} ($(a)!0.75!(b)$) ;
\draw[ thick, -Straight Barb] ($(a)!0.35!(c)$)-- node[right,cyan!70!blue] {\raisebox{-6mm}{$p_4$}} ($(a)!0.75!(c)$) ;
\node[starburst,draw, minimum width=.5em, minimum height=.1em, fill=yellow!50!white,scale=1] at (4.24,.9) {} ;
\draw[ thick, -Straight Barb] ($(d)!0.35!(a)$)-- node[above,red] {\raisebox{0mm}{$p_1$}} ($(d)!0.8!(a)$) ;
\draw[ thick, -Straight Barb] ($(e)!0.35!(a)$)-- node[below,red] {\raisebox{0mm}{$p_2$}} ($(e)!0.8!(a)$) ;
\draw[gray,  Straight Barb-Straight Barb] (4.5,.9)+(0:.5) arc (0:45:.6);
\node[gray] at (5.2,1.2) {$\theta$} ;
\end{tikzpicture}
\end{center}
\vspace{-1em}
\caption{Kinematic constraints on massless $12\rightarrow 34$ scattering as viewed from the celestial sphere. Because of momentum conservation, the scattering process can be embedded in a three-dimensional surface. If we look at how this surface intersects the celestial sphere, we find that there are non-trivial constraints on the operator positions. }
\label{fig:4pt}
\end{figure}
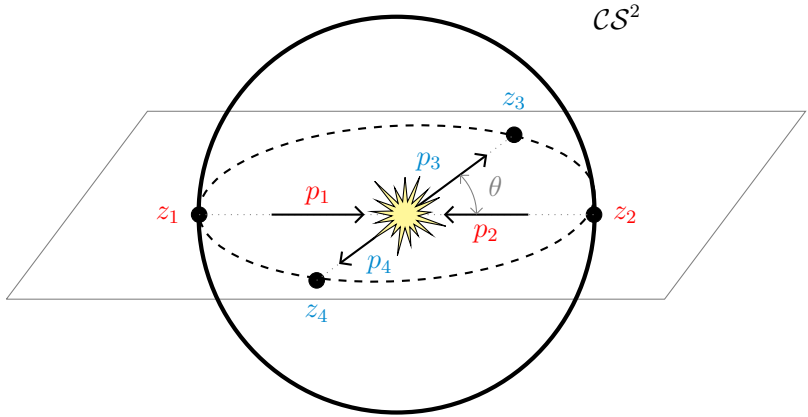

Another fundamental feature is that correlators are supported only on certain well-defined patches of the celestial sphere. Let us illustrate how this \emph{celestial geometry} works for $12 \to 34$ massless scattering understood as a $4$-pt correlator \cite{Pasterski:2017ylz}. As shown in Fig.~\ref{fig:4pt}, momentum conservation restricts the kinematics to a hyperplane in momentum space, and our task is to see how this surface intersects the celestial sphere. Let us start with the momentum space amplitude. In the center-of-mass frame, the scattering process is parameterized by two invariants: the total incoming energy $\sqrt{s}=E$ and the scattering angle $\theta$. Upon performing a rotation we can take the two incoming particles to enter at the north and south poles
        \be
    p_1=E(1,0,0,1),~~~~p_2=E(1,0,0,-1).
    \ee
    By a further rotation around the polar axis we can put the outgoing particles in the $\phi=0$ plane
       \be
    p_3=E(-1, \sin\theta, 0, \cos\theta),~~~~p_4=E(-1,-\sin\theta,0,-\cos\theta).
    \ee
Going to the celestial basis amounts to holding all angles fixed while integrating over the relative energy scales.  In this $12\rightarrow 34$ context this means we are integrating over a family of center-of-mass frames. We also see that for any such $E$ we have an in-out-in-out ordering on the great circle $\phi=0$.   To make this more explicit, let us introduce the cross-ratio $z$ of the vertex operators
\be\label{eq:z}
z = \frac{(z_1 - z_2)(z_3 - z_4)}{(z_1 - z_3)(z_2 - z_4)}.
\ee
Simple geometry tells us that $z = 2/(1 - \cos\theta)$. Reality of $\theta$ implies that the celestial correlator for the $12 \to 34$ scattering has support only on the interval $z \in (1,\infty)$.
 Since this is a Lorentz-invariant statement, the same ordering has to hold in any frame. By such an overall boost/rotation, the great circle considered above can be mapped to any other circle on the celestial sphere.

The story becomes even more interesting when we consider the crossed processes, e.g., $1\bar{4}\to \bar{2}3$ and $1\bar{3} \to \bar{2} 4$, where the bar denotes an anti-particle (particle decays, such as $1 \to \bar{2}34$, are not allowed kinematically for massless particles). Applying the in-out-in-out rule immediately tells us that the two processes only have support on the intervals $z \in (-\infty,0)$ and $(0,1)$ respectively. This picture is an imprint of the fact that scattering amplitudes in different crossing channels do not have overlapping support in the kinematic space, though this is normally phrased in terms of the plane-wave basis.

 At this stage we are faced with two natural, interconnected, problems.  The first question is how the support of CCFT correlators generalizes to higher-multiplicity processes.  The second is how the correlators in different crossing channels with overlapping support are related to one another. In this paper we will tackle both of these questions in turn. In section~\ref{sec:kine} we explore the celestial geometry at $n$-point, identifying the support for each crossing channel and 
demonstrating how the different channels tile
the celestial sphere.
We then use this perspective to revisit how we present the invariant data for celestial correlators in section~\ref{sec:crossing}, focusing in particular on the 4-pt case as a familiar example that can help clarify how crossing symmetry in momentum space manifests itself in CCFT.
 
  Understanding these points becomes vital to the cohesive picture of celestial amplitudes and crossing in both the CFT and amplitudes sense. From a CCFT perspective, our ability to extract symmetries from OPEs relies on going to \emph{complexified} points $z_i$~\cite{Pate:2019lpp,Guevara:2021abz,Strominger:2021lvk,Himwich:2021dau}, making it crucial to understand how to analytically continue from the celestial sphere to a signature where the in/out labels are no longer invariant. From an amplitudes perspective, our ability to go between signatures~\cite{Atanasov:2021oyu,Atanasov:2021cje,Crawley:2021auj,Pasterski:2022lsl} should be intimately connected to the prescriptions for analytic continuations between channels that avoid Landau singularities (see \cite{Mizera:2021ujs,Mizera:2021fap,HM} for recent progress). It would be quite interesting if consistency conditions in CCFT could inform how to prescribe such crossing continuations in the massless case,  though we leave such investigations to future work. Some highlights of the results herein are as follows. 

\paragraph{Selection Rules}
  The  support of $n$-point correlation functions on the celestial sphere is most cleanly phrased in terms of \emph{celestial circles}.
 Namely, scattering in a given crossing channel is disallowed if there exists a circle that separates all the in from all the out punctures.
This geometric picture not only explains the in-out-in-out constraint on the $4$-point function, but also can be applied to a general $n$-point process, where the analytic expressions for the constraints are much more intricate. 
For example, consider what happens when we hold $n{-}1$ punctures fixed and look at the domain of support of the $n$-th puncture for different crossing channels.  At $n=5$ the different crossing channels uniformly tile the celestial sphere once. Starting at $n=6$, the celestial sphere is covered multiple times by scattering amplitudes in different channels, so that there will be a fixed number of channels with support for a given configuration of punctures on the celestial sphere. The degree of this covering $\deg(n)$ is given by
\be
\begin{tabular}{c|c|c|c|c|c|c|c|c|c} 
 $n$ & $5$ & $6$ & $7$ & $8$ & $9$ & $10$ & $11$ & $12$ & $\cdots$ \\ 
 \hline
 $\deg(n)$ & $1$ & $6$ & $22$ & $64$ & $163$ & $382$ & $848$ & $1816$ & $\cdots$ \\ 
\end{tabular}
\ee
and is combinatorially related to the cake-cutting problem in $\R^3$. It grows exponentially with $n$. Codimension-$1$ boundaries of the allowed regions correspond to valid $4$-point processes and the channels on either side of such boundaries have signs of the energies for the remaining $n{-}4$ particles flipped. In other words, the allowed regions are glued together at $(n{-}4)$-fold simultaneous soft limits.

\paragraph{Invariant Data}
As discussed in~\cite{Law:2019glh,Arkani-Hamed:2020gyp}, for $n=4$, any celestial correlator can be expressed in terms of the following invariant data: the cross-ratio $z$ and the sum of the conformal dimensions $\Delta$. Here, we identify a generalization of this statement to $n$-point correlators. For $n \geq 5$, the invariant data can be labeled by $n{-}3$ algebraically-independent complex cross ratios
\be\label{eq:r}
r_{ijkl} = \frac{(z_i - z_j)(z_k - z_l)}{(z_i - z_k)(z_j - z_l)},
\ee
as well as $n$ conformal dimensions, with translation invariance imposing $4$ differential constraints among them. The latter is spelled out in \eqref{eq:sij-constraints}. This analysis gives the total of $3n{-}10$ real degrees of freedom, agreeing with the number of independent Mandelstam invariants for any Poincar\'e-invariant plane-wave amplitude.

\paragraph{Exchange Symmetry} 
In addition to the continuous data accounted for above, the operators come with a discrete label specifying whether the particle is incoming or outgoing. The selection rules for channel support highlight the effect of this label on the celestial correlators. However, this is only part of the story since two channels supported on the same puncture configuration will still probe different regions of phase space. Nonetheless, there is a simpler aspect of crossing that more readily carries over to the celestial basis. Whenever we start with a plane-wave amplitude invariant under the exchange of two particles, the corresponding CCFT correlator will inherit this symmetry. For instance, the $4$-point MHV amplitude of gravitons can be written as
\be
\mathcal{A}_4(1^+ 2^- 3^+ 4^-) \;\propto\; \frac{\langle 24 \rangle^2 [13]^2}{ [24]^2 \langle 13 \rangle^2}
f(s,t)\delta^4(p_1 {+} p_2 {+} p_3 {+} p_4),
\ee
where $f(s,t) = G_{N}\frac{u^3}{st} + \ldots$~. The corresponding CCFT correlator takes the form
\be
\tilde{\mathcal{A}}_4(1^+ 2^- 3^+ 4^-) \;\propto\; \hat{\delta}(\Im\, z) \Theta(\epsilon_i) \frac{z_{24}^2 \bar{z}_{13}^2}{\bar{z}_{24}^2 z_{13}^2} \prod_{i<j} |z_{ij}|^{\frac{\Delta}{3} - \Delta_i - \Delta_j}  g(\Delta,z),
\ee
where $\hat{\delta}(\Im\, z)$ is the crossing-symmetric delta function imposing the support on the celestial circle and $\Theta(\epsilon_i)$ are channel-dependent step-functions implementing the in-out-in-out constraint.
The $1\leftrightarrow 3$ exchange symmetry of the plane-wave amplitude implies crossing symmetry of the CCFT correlator, i.e.,
\be\label{eq:fg}
f(s,t) = f(t,s) \qquad \Leftrightarrow \qquad g(\Delta, z) = g(\Delta, 1{-}z).
\ee
We generalize this statement to $n$-point functions. This exchange symmetry of the CCFT data follows directly from the extrapolate dictionary~\cite{Pasterski:2021dqe,Donnay:2022sdg}, whereby the celestial operators can be expressed as limits of local bulk operators smeared along null generators of the conformal boundary.

\section{Kinematic Constraints on Massless Scattering}\label{sec:kine}

In this section we will consider how translation invariance turns into constraints on the support of celestial amplitudes for generic $n$. The low point $n\le 4$ cases have been examined in~\cite{Pasterski:2017ylz,Law:2019glh}, while the form of the higher point integrand has been investigated in~\cite{Schreiber:2017jsr}. The focus here is to set up the problem in a more geometric manner so that we can understand the indicator functions that appear in those references.

\subsection{Crossing channel support}

Let us start by writing the massless momenta~\eqref{pi} in terms of the in/out label $\epsilon_i = \pm 1$, the energy $\omega_i>0$, and a reference null vector $q_i$
\be\label{qi}
p_i=\epsilon_i \omega_iq_i,\qquad
q_i=\big( 1 + z_i \bar{z}_i, \; z_i + \bar{z}_i, \; -i(z_i - \bar{z}_i),\; 1 - z_i \bar{z}_i \big).
\ee
The vector $\vec{\epsilon}$ of the signs of energies labels a crossing channel (we equate $\vec{\epsilon}$ and $-\vec{\epsilon}$ which are indistinguishable for our purposes). The momentum conserving delta function enforces
\be\label{eq:momem_cons}
\sum_{i=1}^n p_i =\sum_{i=1}^n \epsilon_i \omega_iq_i=0,
\ee
which we can rewrite in the suggestive form
\be\label{qom}
Q \omega :=\Big(\epsilon_1q_1^\mu~\cdots~\epsilon_nq_n^\mu\Big)\left(\begin{array}{c}
     \omega_1  \\ \vdots \\ \omega_n
\end{array}\right)=0.
\ee
The celestial amplitude~\eqref{Acele} integrates over all positive $\omega_i$.  A set of phases and punctures $\{(\epsilon_i, z_i,\bz_i)\}$ is an allowed configuration of celestial operators if the linear equation~\eqref{qom} has a positive solution in $\omega$.  To see when this is the case, we can use the following theorem from~\cite{pjm/1103044952}, whose proof is included in App.~\ref{app:thm}. 

\begin{theorem}[Jackson (2.2) \cite{pjm/1103044952}]\label{thm1}
For an $m\times n$ matrix $Q$, the following are equivalent
\begin{enumerate}[(i)]
    \item $Q{\omega}=0$ has no positive solution ${\omega}\ge0$. 
    \item There exist $v$ such that $v Q>0$.
\end{enumerate}
\end{theorem}

\noindent  For our configuration matrix $Q$ in~\eqref{qom} we need to take $n$ to be the number of external scattering states and $m=4$, matching the bulk spacetime dimension.  We now make the following claim about when a given configuration is disallowed.

\begin{theorem}\label{thm2}
A configuration $\{(\epsilon_i, z_i,\bz_i)\}$ is kinematically disallowed if and only if there is a celestial circle dividing the incoming and outgoing particles.
\end{theorem}

\noindent This follows directly from the proof of Jackson's theorem once we can show that for our class of $Q$'s, the requirement (ii) in Thm.~\ref{thm1} is equivalent to the existence of a celestial circle dividing the incoming and outgoing particles.  We will do this in three steps.  First, we will need to argue that for our type of $Q$ the only $v_\mu$ which can exclude it will be spacelike.  Then we will construct an isomorphism between oriented circles on the celestial sphere and spacelike co-vectors $v_\mu$. Finally, we will show that  $vQ>0$ if this circle divides the incoming and outgoing particles.

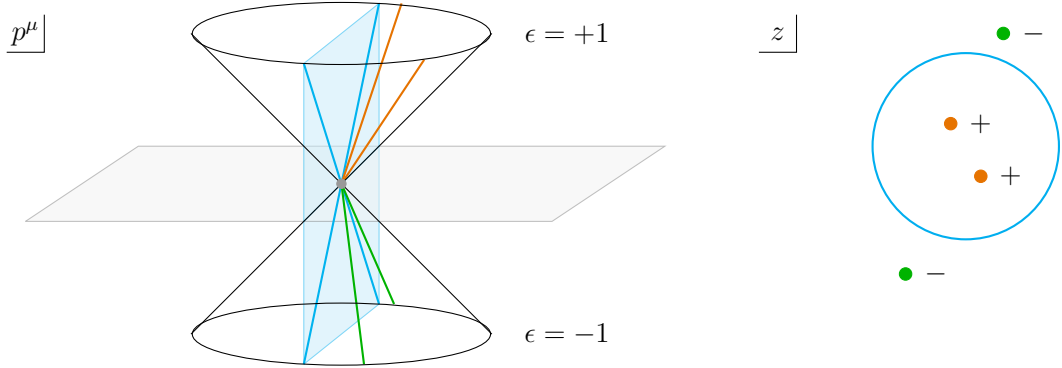
\begin{figure}[t]
\begin{center}
\begin{tikzpicture}[scale=1, every node/.style={scale=.9}]
\draw[cyan,fill=cyan!20!white,opacity=.5] (-.5+4.2,-.4-1) -- (.5+4.2,.4-1) -- (.5+4.2,.5+1) --  (-.5+4.2,-.5+1) --  (-.5+4.2,-.4-1);
\draw[gray,fill=gray!10!white,opacity=.5] (0,0+.5) -- (7,0+.5) -- (8.5,1+.5) -- (1.5,1+.5) -- (0,0+.5);
\draw[cyan,fill=cyan!20!white,opacity=.5] (-.5+4.2,-.5+1) -- (.5+4.2,.5+1) -- (.5+4.2,.4+3) --  (-.5+4.2,-.4+3) --  (-.5+4.2,-.5+1);
\draw[thick,cyan!90!white] (4.2-.5,1-2.4) --(4.2+.5,1+2.4);
\draw[thick,cyan!90!white] (4.2-.5,1+1.6) --(4.2+.5,1-1.6);
\draw[thick,green!70!black] (4.2,1) --(4.2+.7,1-1.6);
\draw[thick,green!70!black] (4.2,1) --(4.2+.3,1-2.4);
\draw[thick,orange!90!black] (4.2,1) --(4.2+.8,1+2.4);
\draw[thick,orange!90!black] (4.2,1) --(4.2+1.1,1+1.65);
\draw[] (4.2,1+2) ellipse (4.8 em and 1em) ;
\draw[] (4.2,1-2) ellipse (4.8 em and 1em) ;
\draw[] (4.2+2,1+2) -- (4.2-2,1-2);
\draw[] (4.2-2,1+2) -- (4.2+2,1-2);
\draw[gray!80!white,fill]  (4.2,1)  circle (.15em);
 \node[] at (7.2,3) {$\epsilon=+1$} ;
  \node[] at (7.2,-1) {$\epsilon=-1$} ;
   \node[] at (0,3) {$p^\mu$} ;
   \draw[] (0-.25,3-.25) -- (0+.25,3-.25) -- (0+.25,3+.25);
    \node[] at (10,3) {$z$} ;
      \draw[] (10-.25,3-.25) -- (10+.25,3-.25) -- (10+.25,3+.25);
  \draw[thick,cyan] (12.5,1.5) circle (3em);
  \draw[orange!90!black,fill]  (12.5+.2,1.5-.4)  circle (.2em) node[right,black] {$~+$};
  \draw[orange!90!black,fill]  (12.5-.2,1.5+.3)  circle (.2em) node[right,black] {$~+$}; 
  \draw[green!70!black,fill]  (12.5+.5,1.5+1.5)  circle (.2em) node[right,black] {$~-$};
  \draw[green!70!black,fill]  (12.5-.8,1.5-1.7)  circle (.2em) node[right,black] {$~-$}; 
  \node[] at (14.5,0) {};
\end{tikzpicture}
\end{center}
\caption{Left: Hyperplanes through the origin of momentum space with spacelike normal (blue) divide the set of on-shell momentum-directions in two (orange and green).  A dual vector $v_\mu$ will satisfy $vQ>0$ if all of the rays are on the same side of this hyperplane. Right: The antipodal identification of the incoming and outgoing celestial spheres implies the corresponding circle divides the incoming (orange) from the outgoing (green) particles. }
\label{fig:planecut}
\end{figure}

While we do not need to attach a Lorentzian metric to $\mathbb{R}^4$ to use Jackson's theorem,  for visualization it is useful to split the the space of co-vectors $v_\mu$ into the three kinds we get in $\mathbb{R}^{1,3}$: spacelike, timelike, or null. The null case can be thought of as a limiting version of either, and will eventually correspond to the limit of shrinking the celestial circle corresponding to a spacelike $v$ down to a point. There is a one-to-one map  between these co-vectors and hyperplanes through the origin in momentum space $\mathbb{R}^{1,3}$. Meanwhile, the columns of $Q$ are vectors lying on the forward or past `lightcones' in this space. The statement that $vQ>0$ means that all of the columns of $Q$ lie on the appropriate side of this hyperplane. As illustrated in Fig.~\ref{fig:planecut}, any spacelike hyperplane (with timelike normal) through the origin in momentum space (grey) will intersect these cones only at the origin. Thus the only configurations they can exclude are the all-in and all-out  scattering processes. However, we will see that these cases can also be excluded by hyperplanes with spacelike normal (blue), so it is sufficient to focus on spacelike $v$'s.  This completes step 1.\footnote{For a spacetime perspective wherein spacelike-normal hyperplanes appear in the definition of currents and their corresponding charges in radially quantized CCFT see~\cite{Pasterski:2022jzc}.}

For step 2 we can write down an explicit isomorphism between celestial circles and spacelike co-vectors $v_\mu$.  Recall that the choice of reference  momenta $q^\mu(z,\bz)$ in~\eqref{qi} amounts to an embedding of the celestial sphere into the canonical section $q^0+q^3=2$ of the forward light cone. If we now let $z=x+iy$, the circle
\be\label{circ}
(x-x_0)^2+(x-y_0)^2=R^2
\ee
can be written as the intersection of this canonical section and a hyperplane through the origin whose normal vector is proportional to
\be\label{vr}
v_\mu=\alpha\left(\tfrac{1}{2}(R^2-1-x_0^2-y_0^2),\, x_0,\, y_0,\, \tfrac{1}{2}(R^2+1-x_0^2-y_0^2)\right).
\ee
Only the sign of $\alpha$ matters from the point of view of condition (ii) and we can use this to assign an orientation to this circle which we will use to distinguish points inside the circle from points outside the circle on the celestial sphere. From the point of view of $\mathbb{R}^{1,3}$, this sign flips us between the two half spaces divided by the corresponding hyperplane.  Noting that~\eqref{vr} sweeps out a hyperboloid of radius $v^2=\alpha^2 R^2>0$ completes step 2.

For step 3 we just need to verify that the configuration where all of the rows of $Q$ are on the same side of the hyperplane corresponding to~\eqref{vr} indeed implies that the incoming and outgoing particles are on opposite sides of the circle~\eqref{circ}.  This is straightforward since
\be
v Q>0~~~\Leftrightarrow~~~ \epsilon_i v_\mu q^\mu_i>0
\ee
for all $i$. Thus, all the outgoing particles will have $v_\mu q^\mu_i>0$ while all the incoming particles will have $v_\mu q^\mu_i<0$, which by our isomorphism between hyperplanes and circles puts them on opposite sides of the celestial circle corresponding to $v$.

\begin{figure}[t]
\centering
\vspace{-0.5em}
\begin{tikzpicture}[scale=2]
\definecolor{darkgreen}{rgb}{.0, 0.5, .1};
\draw[white,fill=gray!90!white,opacity=.5] (-2,-1.5)-- (2,-1.5) -- (2,1.5) -- (-2,1.5) -- (-2,-1.5)  ;
\draw[fill=white](.07,.05) circle[radius=1] ;
\draw[](-.07,-.05) circle[radius=1.02];
\draw[fill=gray!90!white,opacity=.5](-.07,-.05) circle[radius=1.02] ;
\draw[thick](0,0) circle[radius=1] ;
\fill[](0,0) ++(-90:1) circle[radius=1pt] node[below] {$-$};
\fill[](0,0) ++(0:1) circle[radius=1pt] node[right] {$+$};
\fill[](0,0) ++(45:1) circle[radius=1pt] node[above right] {$+$};
\draw[->, thick] (0,-1) arc[radius=1, start angle=-90, end angle=-45];
\end{tikzpicture}
\caption{A geometric route to the in-out-in-out ordering and celestial circle support of allowed $2\rightarrow2$ scattering processes. After placing punctures with $(\epsilon_1,\epsilon_2,\epsilon_3)=(+,+,-)$, the additional $-$ puncture cannot be placed in any of the grey regions by Thm.~\ref{thm2}. Translating and dilating the circles leaves the only allowed position on the arc between the two $+$ punctures.}
\label{fig:4ptdis}
\end{figure}
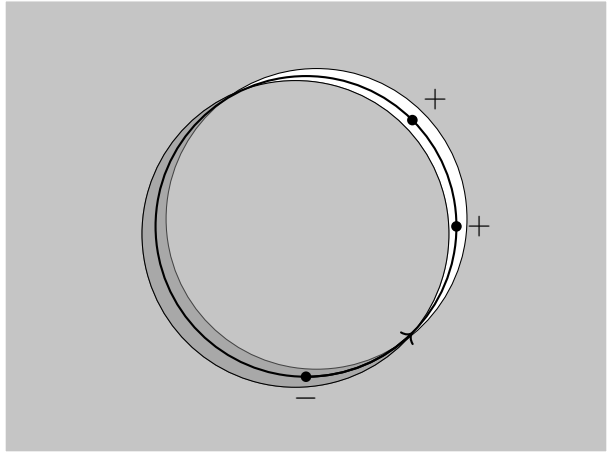

\paragraph{Revisiting low-point kinematics} Let us now use this theorem to re-derive the familiar constraints on celestial correlators~\cite{Pasterski:2017ylz}. First of all, for any $n$, Thm.~\ref{thm2} immediately excludes any $0 \leftrightarrow n$ and $1 \leftrightarrow n{-}1$ process, because in either case one can simply draw a celestial circle that separates all the in from out states, unless two of them are collinear.\footnote{This caveat allows for the contact term 2-point function we get from Mellin transforming the familiar momentum space inner product from the single particle Hilbert space, corresponding to a $1\to 1$ process.} Returning to the remaining $2 \to 2$ processes at $n=4$, we expect the punctures to be restricted to a circle by momentum conservation. The in-in-out-out ordering of the punctures on the circle is easily excluded, which leaves us with in-out-in-out as the only valid option. 

Even if we did not know that all four punctures have to be aligned on a circle, we could have arrived at this result as follows. For any $3$ points on the celestial sphere we can draw a circle through them. Consider the circle through two $+$ and one $-$ puncture illustrated in Fig.~\ref{fig:4ptdis}, as well as two deformations of this circle designed so that the $+$ and $-$ particles are on opposite sides. By a Lorentz transformation we can map any three punctures to points of our choice, so this drawing is generic for non-degenerate configurations.  In the case where we take these deformations to be infinitesimal these two circles exclude all but the arc between the two $+$ punctures for the position of the fourth point  which is a $-$ puncture.  This expediently reproduces both the celestial circle and the in-out-in-out ordering we reviewed in the introduction.\footnote{This story can be generalized to $(2,2)$ signature, in which case the celestial circles are replaced by corresponding celestial hyperbolae.}

\subsection{Tiled covering of the celestial sphere}\label{sec:tiled}

In the interest of understanding what happens when we analytically continue the position of one of the external operators, we will now turn to the following question:
\vspace{1em}

\noindent {\it For a given crossing channel $\vec{\epsilon}$ and the positions of punctures $i=1,2,\ldots,n{-}1$ fixed, what is the region of support for puncture $n$?}
\vspace{1em}

\noindent In the four-point case this is given by the unique arc on the circle through punctures $\{1,2,3\}$ that obeys the in-out-in-out ordering.  We claim that, more generally, the excluded region of the celestial sphere has the following properties:
\begin{enumerate}
    \item  The disallowed region can be written as union of a finite number of disks whose boundaries are circles through 3 of the fixed punctures.
    \item The arcs that make up the boundary of this union corresponds to allowed $2\rightarrow2$ scattering processes.
\end{enumerate}

\noindent
Since the channels $\vec{\epsilon}$ and $-\vec{\epsilon}$ have the same support, without loss of generality we can set the final particle to be outgoing, i.e., $\epsilon_n = -1$. 
We note that if the $n{-}1$ particle configuration is allowed, there is no restriction on the placement of the $n$-th puncture. We can see this by the fact that Thm.~\ref{thm2} tells us there is no circle separating the in particles from the out particles within the $n{-}1$ particle process. By contrast, if the $n{-}1$ particle configuration is not allowed, we know that  we can draw a celestial circle that divides the in particles from the out particles. If the $n$-th particle is placed anywhere on the side of circle with the $\epsilon_i=-1$ punctures, the corresponding $n$-particle configuration will be disallowed too.  

\begin{figure}[t]
\centering
\vspace{-0.5em}
\begin{tikzpicture}[scale=2]
\definecolor{darkgreen}{rgb}{.0, 0.5, .1};
\draw[thick,fill=white!90!gray](0,0) circle[radius=1] ;
\fill[](0,0) ++(-90:1) circle[radius=1pt] node[below] {$j$};
\fill[](0,0) ++(0:1) circle[radius=1pt] node[right] {$k$};
\fill[](0,0) ++(45:1) circle[radius=1pt] node[above right] {$l$};
\draw[->, thick] (0,-1) arc[radius=1, start angle=-90, end angle=-45];
\node[] at (0,0) {$\boldsymbol{-}$};
\node[] at (2,0) {$\boldsymbol{+}$};
\node[] at (-2,0) {};
\end{tikzpicture}
\caption{Sign of $\mathrm{Im}(r_{ijkl})$ as determined by the position of particle $i$ given the oriented circle connecting particles $(jkl)$.}
\label{4ptcirc}
\end{figure}
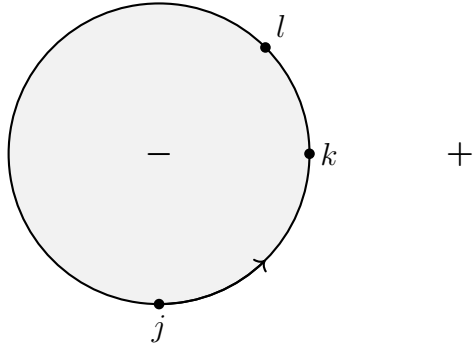

What, then, are the allowed $n$-particle configurations? Since we want to attach an orientation that distinguishes the inside from the outside, we will label such circles with a cyclically-ordered triplet $(jkl)$ which  defines a hyperplane  via
\be
v_\mu^{(jkl)}=\varepsilon_{\mu\nu\sigma\rho} \epsilon_j q_j^\nu \epsilon_k q_k^\sigma \epsilon_l q_l^\rho,
\ee
where $\varepsilon_{\mu\nu\sigma\rho}$ is the totally-antisymmetric tensor.
For later reference, we note that the dot product with one of the columns of $Q$
\be\label{eq:4x4minors}
v_\mu^{(jkl)} \epsilon_i q_i^\mu= \det Q_{ijkl}
=\epsilon_i\epsilon_j\epsilon_k\epsilon_l |z_{ik}z_{jl}|^2\, \mathrm{Im} (r_{ijkl})
\ee
is the corresponding  $4\times4$ minor of $Q$ whose sign is determined by the crossing channel and the imaginary part of the cross ratio~\eqref{eq:r}
\be
r_{ijkl}=\frac{z_{ij}z_{kl}}{z_{ik}z_{jl}},
\ee
which flips when we cross the circle, as illustrated in Fig.~\ref{4ptcirc}. In what follows we will consider the situation where the first $n{-}1$ punctures are at generic positions. In this case only triplets will be co-circular. 
Phrased more mathematically, the moduli space for the `generic' $n$-point celestial correlators we are considering is
\be\label{eq:moduli}
\left[ (\CP^1)^n - \bigcup_{i,j,k,l} \{ \Im\, r_{ijkl} = 0 \} \right] /\, \SL(2,\C)
\ee
for $n \geq 5$, where the action of $\SL(2,\C)$ allows us to freeze three punctures. It seems to be a non-linear cousin of the moduli space ${\cal M}_{0,n}$ familiar from string perturbation theory, which can be obtained by removing ``$\Im$'' from \eqref{eq:moduli}. It would be fascinating to further explore combinatorial and topological aspects of this moduli space, such as its compactification.

Now let us start with one of the circles separating the in from the out particles among the first $n{-}1$ fixed punctures. We will refer to the interior as the side which contains the $\epsilon_i=-1$ punctures.  For any such circle on the celestial sphere we will be able to continuously deform it to some subset of these $(jkl)$ without crossing any of the fixed punctures. If particle $n$ is in any of these deformed circles, the configuration is again disallowed.  We thus confirm property 1.

The circles defining the boundary of this region correspond to extreme rays of the cone satisfying $vQ_{i}>0$ for all $i<n$.  The interior of this cone is open.  Correspondingly, we see by  our construction that all the $\epsilon_i=-1$ punctures are in the interior of the disallowed region for puncture $n$. Meanwhile the $\epsilon_i=+1$ punctures are either in or on the boundary of the allowed region. Combined with property 1, the boundary of this region is thus a union of circular arcs through two $\epsilon_i=+1$ punctures. Let $j$ and $k$ refer to an adjacent pair of punctures on this boundary. If the third puncture $l$ defining this circle had the same sign $\epsilon_l=+1$, it would circumscribe the allowed region by construction, since a circle that intersects the boundary transversely cannot correspond to a vector inside or on this cone. We can do a continuous deformation of this circle on the Riemann sphere that keeps points $j$ and $k$ fixed and moves away from the third point $l$ towards the $\epsilon_i=-1$ punctures (any element of this family can be infinitesimally deformed to a circle satisfying Thm.~\ref{thm2}.).  The arc between $j$ and $k$ on the original circle will be in the interior of region disallowed by the deformed circle and thus cannot be a boundary of the disallowed region.  We thus see that the boundary is a union of $(+,+)$ arcs on an $(-,+,+)$ circle, demonstrating property 2. 

\vspace{1em}

\begin{figure}[ht!]
\centering
\includegraphics[scale=1,valign=c]{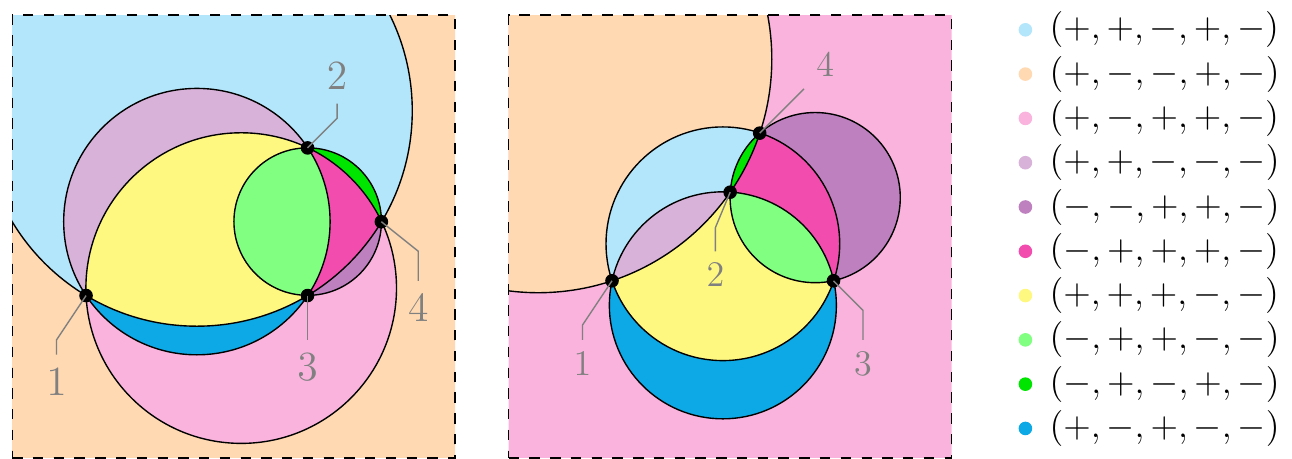}
\vspace{1em}
\caption{Tiling of the celestial sphere for $n=5$, holding $\{z_1,z_2,z_3,z_4\}$ fixed in two different puncture configurations, together with the corresponding channel labels $\vec{\epsilon}/\mathbb{Z}_2$
}
\label{5pttile}
\end{figure}

\noindent So far we have kept the channel fixed. Let us now generalize to the case where we keep the $n{-}1$ puncture locations fixed and look at all crossing channels. In the 4-point case we saw that different channels tiled the celestial circle. Here we will see that the different crossing channels tile a covering of the celestial sphere.  The degree of this covering, i.e., the number of crossing channels allowed for a given point on the sphere, as a function of $n$ is related to the so-called cake numbers
\be\label{eq:deg}
\mathrm{deg}(n)=2^{n-1}-\mathrm{cake}(n-1)
\ee
where 
\be
\mathrm{cake}(n-1)=\binom{n}{3}+n=\binom{n-1}{0}+\binom{n-1}{1}+\binom{n-1}{2}+\binom{n-1}{3}.
\ee
This grows rapidly with $n$.  For $n=1,2,3,...$ the degree is
\be
0,0,0,0,1,6,22,64,163,382,848,1816,\ldots
\ee
for generic puncture configurations. In particular the 5-point channels tile the celestial sphere exactly once. This is illustrated for two different puncture configurations in Fig.~\ref{5pttile}.

A priori there are $2^{n-1}$ possible assignments of $\vec{\epsilon}/\mathbb{Z}_2$.  To verify~\eqref{eq:deg},  we want to show that the number of disallowed channels over a given point is the cake number cake$(n-1)$. Now the cake number $\mathrm{cake}(n-1)$ is the maximum number of pieces of cake you can get by cutting a 3D cake with $n-1$ planes. In order to prove~\eqref{eq:deg} we will relate the question we are after to a problem that is amenable to a generalization of the standard a 3D cake cutting problem's proof.

By Thm.~\ref{thm1}, the $n$ particle configuration is disallowed if there is a $v$ such that $vQ>0$.  In the same manner that the $v_\mu$ define hyperplanes in momentum space, the $q^\mu_i$ cut up the dual momentum space into chambers. For every such chamber $v\in V$ the entries $\nu_i=\mathrm{sgn}(v_\mu q_i^\mu)$ have definite sign.  Taking $\epsilon_i=\nu_i$ gives a configuration that is disallowed for each chamber.  Among the pairs $\{V,-V\}$ one of them will have $\epsilon_n=-1$ matching our choice for reducing the $\mathbb{Z}_2$ redundancy.  We thus want to count the number of pairs of chambers $\{V,-V\}$. 

We can now turn this into a 3D cake cutting problem by restricting to an affine subspace which will hit only one of the pair $\{V,-V\}$.  We need to take care with this choice.  For instance the subspace $v_0=1$ will intersect both chambers when $n=1$.  Instead consider the 3D slice $v_\mu q_1^\mu=1$. This lies strictly on one side of the first hyperplane. The remaining $n-1$ hyperplanes will cut up this volume, which will only hit one of the two chambers $\{V,-V\}$ since the sign of the first entry is fixed. While the hyperplanes defined by the $q_i$ are not generic, their intersections with the $v_\mu q_1^\mu=1$ hypersurface still satisfy the necessary criteria for the number of `slices' to be cake$(n-1)$. Namely, the fact within this 3-volume no two planes are parallel, no two of their intersection lines are parallel, and there is no point in common to four or more planes follows from the nature of the 4D uplift and we can use the standard proof~\cite{OEIS}.

\pagebreak

Finally, we note that some of these channels will cover a full copy of the Riemann sphere. Because a channel will have full support for the $n$-th puncture only when the $n-1$ configuration is allowed we have
\be
\mathrm{full}(n)=2\mathrm{deg}(n-1)
\ee
corresponding to the two relative signs of $\vec{\epsilon}_{i<n}$.

\begin{figure}[bt]
\begin{center}
\begin{tikzpicture}[scale=.7, every node/.style={scale=.9}]
\draw[thick,fill=cyan!30!white] (0,0) -- (7,0) -- (8.5,2) -- (1.5,2) -- (0,0);
\draw[thick,fill=orange!30!white] (0,0-2.5) -- (7,0-2.5) -- (8.5,2-2.5) -- (1.5,2-2.5) -- (0,0-2.5);
\draw[thick,fill=magenta!30!white] (0,0-5) -- (7,0-5) -- (8.5,2-5) -- (1.5,2-5) -- (0,0-5);
\node[] at (-1.5,1) {$\mathbb{CP}^1$};
\node[] at (-1.5,1-2.5) {$\mathbb{CP}^1$};
\node[] at (-1.5,1-5) {$\mathbb{CP}^1$};
\node[] at (-1.5,1-2.5/2) {$\times$};
\node[] at (-1.5,1-2.5/2*3) {$\times$};
\node[] at (-1.5,1-2.5/2*5) {$\vdots$};
\coordinate (a) at (4.35-2.1+.02,.9+.08) ;
\coordinate (b) at (3.7,1.5) ;
\coordinate (c) at (2.9,.75-.1);
\coordinate (d) at (5.2,.8);
\coordinate (e) at (6.2,1.2);
\coordinate (a2) at ($(a)+(0,-2.5)$);
\coordinate (b2) at ($(b)+(0,-2.5)$);
\coordinate (c2) at ($(c)+(0,-2.5)$);
\coordinate (d2) at ($(d)+(0,-2.5)$);
\coordinate (e2) at ($(e)+(0,-2.5)$);
\draw[fill=violet!30!white] (2.8,2-2.5)  to [out=210,in=100]  ($(a)+(0,-2.5)$) to [out=-20,in=190]  ($(c)+(0,-2.5)$)  to [out=-10,in=190]  ($(d)+(0,-2.5)$)  to [out=40,in=-90] (5.9,2-2.5) -- (2.8,2-2.5);
\draw[fill=green!90!black] ($(a)+(0,-2.5)$) to [out=-20,in=190]   ($(d)+(0,-2.5)$)   to [out=210,in=-60]  ($(a)+(0,-2.5)$) ;
\draw[fill=violet!50!white] (2.3,2-5)  to [out=220,in=120]  ($(a)+(0,-5)$) to [out=40,in=180]  ($(b)+(0,-5)$)  to [out=20,in=150]  ($(e)+(0,-5)$)  to [out=20,in=-120] (7.5,2-5) -- (2.2,2-5);
\draw[fill=green!70!white]  ($(b)+(0,-5)$)  to [out=20,in=150]  ($(e)+(0,-5)$)  to [out=20,in=0] ($(d)+(0,-5)$) to [out = 130, in=-20] ($(b)+(0,-5)$)  ;
\draw[fill=yellow!70!white]   ($(d)+(0,-5)$) to [out = 130, in=-20] ($(b)+(0,-5)$) to [out = 200, in=80] ($(c)+(0,-5)$) to [out = -20, in= 200]  ($(d)+(0,-5)$)  ;
\draw[fill=magenta!70!white]  ($(a)+(0,-5)$) to [out=40,in=180]   ($(b)+(0,-5)$) to [out = 200, in=80]   ($(c)+(0,-5)$) to [out = 170, in=-20]  ($(a)+(0,-5)$) ;
\draw[fill=cyan!95!black]  ($(c)+(0,-5)$)  to [out = -20, in= 200]  ($(d)+(0,-5)$) to [out=0,in=20]  ($(e)+(0,-5)$) to [out = 230, in=-30]  ($(c)+(0,-5)$) ;
\draw[gray] ($(a)+(0,.5)$) -- ($(a)+(0,-7.5)$) node[below] {$1$};
\draw[gray] ($(b)+(0,.5)$) -- ($(b)+(0,-7.5)$) node[below] {$3$};
\draw[gray] ($(c)+(0,.5)$) -- ($(c)+(0,-7.5)$) node[below] {$2$};
\draw[gray] ($(d)+(0,.5)$) -- ($(d)+(0,-7.5)$) node[below] {$4$};
\draw[gray] ($(e)+(0,.5)$) -- ($(e)+(0,-7.5)$) node[below] {$5$};
\draw[fill] (a) circle (.2em) ;
\draw[fill] (b) circle (.2em);
\draw[fill] (c) circle (.2em) ;
\draw[fill] (d) circle (.2em) ;
\draw[fill] (e) circle (.2em) ;
\draw[fill] ($(a)+(0,-2.5)$) circle (.2em) ;
\draw[fill] ($(b)+(0,-2.5)$) circle (.2em);
\draw[fill] ($(c)+(0,-2.5)$) circle (.2em) ;
\draw[fill] ($(d)+(0,-2.5)$) circle (.2em) ;
\draw[fill] ($(e)+(0,-2.5)$) circle (.2em) ;
\draw[fill] ($(a)+(0,-5)$) circle (.2em) ;
\draw[fill] ($(b)+(0,-5)$) circle (.2em);
\draw[fill] ($(c)+(0,-5)$) circle (.2em) ;
\draw[fill] ($(d)+(0,-5)$) circle (.2em) ;
\draw[fill] ($(e)+(0,-5)$) circle (.2em) ;
\node[] at (9,0) {};
\draw[fill,cyan!30!white] (10,2*.8) circle (.2em) node[right,black] {$~(-,+,+,-,+,-)$};
\draw[fill,orange!30!white] (10,1*.8) circle (.2em) node[right,black] {$~(+,-,-,+,+,-)$};
\draw[fill,magenta!30!white] (10,0) circle (.2em) node[right,black] {$~(+,+,-,-,+,-)$};
\draw[fill,violet!30!white] (10,-1*.8) circle (.2em) node[right,black] {$~(+,-,+,+,-,-)$};
\draw[fill,violet!50!white] (10,-2*.8) circle (.2em) node[right,black] {$~(+,-,+,-,+,-)$};
\draw[fill,magenta!70!white] (10,-3*.8) circle (.2em) node[right,black] {$~(+,+,+,-,-,-)$};
\draw[fill,yellow!50!white] (10,-4*.8) circle (.2em) node[right,black] {$~(-,+,+,+,-,-)$};
\draw[fill,green!50!white] (10,-5*.8) circle (.2em) node[right,black] {$~(-,-,+,+,+,-)$};
\draw[fill,green!90!black] (10,-6*.8) circle (.2em) node[right,black] {$~(+,+,-,+,-,-)$};
\draw[fill,cyan!95!black] (10,-7*.8) circle (.2em) node[right,black] {$~(-,+,-,+,+,-)$};
\end{tikzpicture}
\end{center}
\caption{Tiling of the celestial sphere for scattering with $n=6$ holding $\{z_1,z_2,\ldots,z_5\}$ fixed. The support of the various $3\rightarrow 3$ channels are shown, along with the corresponding $\vec{\epsilon}/\mathbb{Z}_2$.}
\label{fig:tiling}
\end{figure}
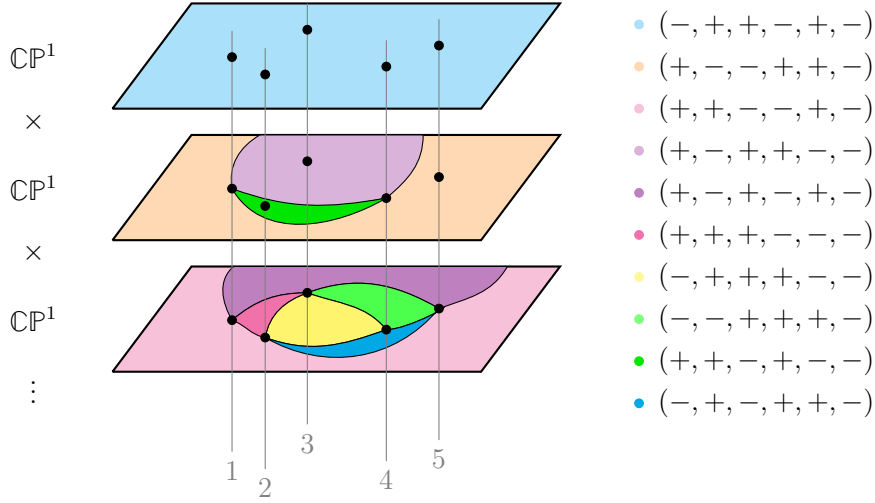

The remaining channels will tile together into copies of $\mathbb{CP}^1$. From our investigation at the beginning of this section, we know that the boundaries of each channel correspond to allowed $2\rightarrow 2$ processes. We can thus attempt to tile together these channels, gluing across boundaries corresponding to $(n-4)$-fold multi-soft limits. The $n=6$ case is is illustrated in Fig.~\ref{fig:tiling}.   The channels on either side of a given boundary arc flips the signs of all but these 4 particles whose punctures define that arc. Because this operation maps $3\rightarrow 3$ and $2\leftrightarrow 4$ properties amongst themselves we have restricted to the former in the figure. We see that there are three disconnected components of this cover. We can move around within each region without hitting a soft limit of scattering. While we need to understand multi-soft limits in order to cross between regions within a fixed $\mathbb{CP}^1$, these region boundaries are different within each sheet of the cover.  As such, being able to relate amplitudes in the channels supported over a fixed configuration on the sphere would inform a prescription for how to approach these multi-soft limits from either side. We will turn to the topic of such channel dependence next.

\section{Invariant Data and Crossing 
}\label{sec:crossing}

We will now turn to the analytic properties of CCFT correlators for general $n$. An important first step is to strip off the kinematics from the dynamics. Recall that a Lorentz transformation implements the following map on the momentum space data
\be
z_i \to z_i' = \frac{az_i + b}{c z_i + d},~~~\omega_i \to \omega_i' = |c z_i + d|^2 \omega_i
\ee
while the crossing channel label $\epsilon_i$ is a Lorentz invariant in $(1,3)$ signature.  The $n$-point amplitude is Lorentz covariant, not invariant. If we start in the plane wave basis, the little group transformation properties of massless single particle states implies that the amplitude for $n$ massless particles with helicity $J_i$ transforms as follows
	\be
	\A_n(\epsilon_i, \omega_i', z_i', \bar{z}_i') = \prod_{i=1}^n \left(\frac{cz_i +d}{\bc \bz_i + \bd}\right)^{\!\!2J_i} \A_n(\epsilon_i, \omega_i, z_i, \bar{z}_i).
	\ee
	Noting that $z_{ij}' = \frac{z_{ij}}{(c z_i + d)(c z_j + d)}$, we can strip off a kinematical factor that captures the little group covariance
	\be\label{caln}
	\A_n(\epsilon_i, \omega_i, z_i, \bar{z}_i) = \prod_{i<j} \left( \frac{z_{ij}}{\bar{z}_{ij}} \right)^{\frac{1}{n-2}(\frac{J}{n-1} - J_i - J_j)} \delta^4(\textstyle\sum_{i=1}^{n}p_i^\mu)\, A_n(s_{ij})
	\ee
	leaving us with a Lorentz invariant amplitude $A_n(s_{ij})$. Here $J = \sum_{i=1}^{n} J_i$ is the total helicity, $s_{ij} = (p_i+p_j)^2$ are the Mandelstam invariants, and translation invariance  implies a distributional support on the locus of momentum conservation, which we've stripped off as well. The dynamics is encoded in the dependence of $A_n$ on the Mandelstam variables. These $s_{ij}$ are free variables for $n\leq 5$, but more generally are constrained by Gram determinant conditions, whereby every $5\times 5$ minor of $s_{ij}$ (treated as a matrix) must vanish.
	
	The CCFT correlator is defined as a Mellin transform in the energy variables $\omega_i$ as in~\eqref{Acele}. This has the effect of diagonalizing boosts in the direction of the four-momentum, so that 
	under $\SL(2,\C)$ this object transforms as follows
	\be
	\tilde{\A}_n(\epsilon_i, \Delta_i, z_i', \bar{z}_i') = \prod_{i=1}^{n} (c z_i + d)^{2h_i} ({\bc \bz_i + \bd})^{2\tilde{h}_i} \tilde{\A}_n(\epsilon_i, \Delta_i, z_i, \bar{z}_i)
	\ee
	where $h_i = \frac{1}{2}(\Delta_i + J_i)$ and $\tilde{h}_i = \frac{1}{2}(\Delta_i - J_i)$. From our discussion in the previous section, we saw that this amplitude has restricted support as a function of the cross ratios $r_{ijkl}$ that depends on the channel $\vec{\epsilon}$. Together with the SL$(2,\mathbb{C})$ transformations, this implies that we can write
		\be\label{eq:Atilde}
	\tilde{\A}_n(\epsilon_i, \Delta_i, z_i, \bar{z}_i) = \Theta_n(\epsilon_i,r_{ijkl}) \prod_{i<j}\left(z_{ij}^{\frac{2}{n-2}(\frac{h}{n-1} - h_i -h_j)} \bar{z}_{ij}^{\frac{2}{n-2}(\frac{\tilde{h}}{n-1} - \tilde{h}_i - \tilde{h}_j)} \right) \tilde{A}_n(\epsilon_i,\Delta_i,r_{ijkl}),
	\ee
	where $h = \sum_{i=1}^{n} h_i$ and $\tilde{h} = \sum_{i=1}^{n} \tilde{h}_i$ and the dynamics is encoded in the Lorentz invariant amplitude $\tilde{A}_n(\epsilon_i,\Delta_i,r_{ijkl})$. 	We can cleanly extract this invariant amplitude as follows. Starting from the spinor helicity products
	\be
	\langle ij\rangle =-2\epsilon_i\epsilon_j \sqrt{\omega_i\omega_j}z_{ij},~~~[ij]=2\sqrt{\omega_i\omega_j}\bz_{ij}
	\ee
we recognize the spin dependent factor in~\eqref{caln} as the following ratio of the angle and square spinor products
\be
\prod_{i<j} \left( \frac{z_{ij}}{\bar{z}_{ij}} \right)^{\frac{1}{n-2}(\frac{J}{n-1} - J_i - J_j)}=\prod_{i<j} \left(\epsilon_i\epsilon_j \frac{\langle ij\rangle}{[ij]} \right)^{\frac{1}{n-2}(\frac{J}{n-1} - J_i - J_j)}.
\ee
We can do the same thing for the $\Delta_i$ dependence by introducing appropriate factors of 
\be
	s_{ij}=\langle ij\rangle[ij] = -4\epsilon_i\epsilon_j \omega_i\omega_j|z_{ij}|^2.
	\ee
This leads us to the compact expression\footnote{This form has a couple of nice features.  First, we see that the choice $\mathcal{N}_{\Delta_i,\epsilon_i}=1$ in~\eqref{Acele} is equivalent to using an integration kernel that involves the crossing invariant combinations $\epsilon_i\epsilon_j\langle ij\rangle$. Moreover, this form is more amenable to changing signature. Namely, we can reintroduce the little group phase as in~\cite{Brandhuber:2021nez} so that the spinor helicity products become
	\be
	\langle ij\rangle = -2\epsilon_i\epsilon_ju_i u_jz_{ij},~~~[ij] = 2\bar u_i\bar u_j\bz_{ij}
	\ee
while replacing each $\d\log\omega_i$ with
	\be
\int\frac{\d u_i}{u_i}\wedge \frac{\d\tilde{u}_i}{{\tilde u}_i},~~~u_i=\sqrt{\omega_i}e^{i\theta_i}.
\ee
In $\mathbb{R}^{1,3}$ the contour is taken by setting $\tilde{u}_i = \bar{u}_i$ (complex conjugate of $u_i$) and integrating over the complex plane.
Changing to Klein space $\mathbb{R}^{2,2}$ amounts to replacing the contour for $\tilde{u}_i$ by treating it as an independent real variable.  
}
\be\label{compact}\scalemath{1}{
\Theta_n\tilde{A}_n
= \int \prod_{i=1}^{n} \frac{\d\omega_i}{\omega_i} \prod_{i<j}\left(\left(-\frac{\epsilon_i\epsilon_j}{2}\langle ij\rangle\right)^{-\frac{2}{n-2}(\frac{h}{n-1} - h_i -h_j)} \left(\frac{1}{2}[ij]\right)^{-\frac{2}{n-2}(\frac{\tilde{h}}{n-1} - \tilde{h}_i - \tilde{h}_j)} \right)\mathcal{A}_n},
\ee
where $\mathcal{A}_n$ is the full momentum-conservation included amplitude.  The factors of $\omega_i^{\Delta_i}$ in the Mellin transform come from the $s_{ij}$-dependence of this kernel. Meanwhile the remaining $\d\log \omega_i$ measure for each particle is the natural scale-invariant measure on $\mathbb{R}_+$. The goal of this section is to establish a good basis for the invariant data $\Delta_i$, $r_{ijkl}$ describing a general correlator. We start with the well-understood special case of $n=4$ \cite{Pasterski:2017ylz,Law:2019glh,Arkani-Hamed:2020gyp,Chang:2021wvv}, in order to clarify some basic points and set stage for the case of general $n$.

	\subsection{\label{sec:four}Four particles}
	
	For $n=4$ we can write the celestial amplitude as follows
	\be
	\tilde{\A}_4(\epsilon_i, \Delta_i, z_i, \bar{z}_i) = \prod_{i<j} \left( \frac{z_{ij}}{\bar{z}_{ij}} \right)^{\!\frac{1}{2}(\frac{J}{3} - J_i - J_j)}  \left(\prod_{i=1}^{4} \int_0^\infty  \d \omega_i\, \omega_i^{\Delta_i-1}\right) \delta^4(\textstyle\sum_{i=1}^{4}p_i^\mu) A_4(s,t).
	\ee
	As discussed in the introduction, a special feature of scattering at 4-point is that it has codimension-1 support on the celestial sphere due to the fact that four $p_i$'s satisfying the momentum conservation constraint cannot themselves span the full four-dimensional momentum space.   The way this manifests itself at the level of how one evaluates the $n=4$ correlator is that $\delta^4(\sum_{i=1}^{4} p_i^\mu)$ cannot be used to localize all four $\omega_i$'s. Instead, the best we can do is to localize three of the energy variables, writing
	\be
	\delta^4({\textstyle\sum}_{i=1}^{4} p_i^\mu ) = \frac{1}{2\omega_4 |z_{13}z_{24}|^2}\, \delta(\Im z) \prod_{i=1}^{3} \delta(\omega_i - \omega_i^\ast)
	\ee
	with
	\be\label{omegastar}
	\omega_1^\ast = - \frac{\epsilon_1 \epsilon_4 \omega_4}{z} \left| \frac{z_{34}}{z_{13}}\right|^2,\quad \omega_2^\ast =  \frac{\epsilon_2 \epsilon_4 \omega_4 (1{-}z)}{z} \left|\frac{z_{34}}{z_{23}}\right|^2, \quad \omega_3^\ast = \epsilon_3 \epsilon_4 \omega_4 (z{-}1) \left|\frac{z_{24}}{z_{23}}\right|^2.
	\ee
	The constraint $\Im z = 0$ on the cross-ratio~\eqref{eq:z}
	\be
	z = \frac{z_{12} z_{34}}{z_{13} z_{24}}
	\ee
	means the four punctures have to be aligned along a great circle. The constraint $\delta(\Im z)$ survives the Mellin transform.  Meanwhile the energy delta functions $\delta(\omega_i - \omega_i^\ast)$ are only saturated when their arguments are positive $\omega_i^\ast > 0$. This puts an additional constraint on $z$ as a function of the channel labels $\epsilon_i$. Namely,
	\be\label{eq:4pt-channels}
	\Theta_4(\epsilon_i,z) = \begin{dcases}
		\theta(z{-}1) \qquad &\text{for}\quad \vec{\epsilon}/\Z_2=(+,+,-,-)\quad [s\mathrm{-channel}],\\
		\theta(-z) \qquad &\text{for}\quad \vec{\epsilon}/\Z_2=(-,+,+,-) \quad [t\mathrm{-channel}],\\
		\theta(z)\theta(1{-}z) \qquad &\text{for}\quad \vec{\epsilon}/\Z_2=(+,-,+,-)\quad [u\mathrm{-channel}],
	\end{dcases}
	\ee
	where $\theta(z)$ is the Heaviside step function. This is the only source of kinematic constraints depending on the crossing channel.
	
	Putting this together we have
	\be\badat{3}\label{A4rough}
	\tilde{\cal A}(\epsilon_i, \Delta_i, z_i, \bar{z}_i) &= \frac{\Theta_4(\epsilon_i,z) \delta(\Im z)}{2 |z_{13}z_{24}|^2} \prod_{i<j} \left(\frac{z_{ij}}{\bar{z}_{ij}} \right)^{\!\frac{1}{2}(\frac{J}{3} - J_i - J_j)} \prod_{i=1}^{3} \left(\frac{\omega_i^\ast}{\omega_4}\right)^{\Delta_i - 1} \\
	&~~~\times \int_0^\infty \mathrm{d}\omega_4\, \omega_4^{\Delta-5} A_4( s^\ast, t^\ast ).
	\eadat\ee
	where $\Delta = \sum_{i=1}^{4} \Delta_i$ and the Mandelstam invariants $(s^\ast, t^\ast, u^\ast)$ evaluated on the support of the delta functions read
	\be
	(s^\ast, t^\ast, u^\ast) = 4 (z-1)\omega_4^2 \left| \frac{z_{24} z_{34}}{z_{23}} \right|^2 \left( 1,\; \frac{1-z}{z},\; -\frac{1}{z} \right).
	\ee
	Note that all the explicit $\epsilon_i$-dependence has dropped out. As a cross check, it is easy to check momentum conservation, $s^\ast + t^\ast + u^\ast = 0$, and that each channel selects correct signs for the Mandelstam invariants, e.g., $s^\ast>0$, $t^\ast,u^\ast<0$ in the $s$-channel with $z>1$.
	
	At this stage we would like to put the expression~\eqref{A4rough} into the form \eqref{eq:Atilde} with the overall $\SL(2,\C)$-covariance factored out. To this end we first note that
	\be\label{omegastar2}
	\frac{\omega_1^\ast}{\omega_4} = \frac{1}{|z|} \left| \frac{z_{34}}{z_{13}}\right|^2,\qquad \frac{\omega_2^\ast}{\omega_4} =  \frac{ |1{-}z|}{|z|} \left|\frac{z_{34}}{z_{23}}\right|^2, \qquad \frac{\omega_3^\ast}{\omega_4} = |1{-}z| \left|\frac{z_{24}}{z_{23}}\right|^2.
	\ee
	is valid in every crossing channel. Moreover, we can rescale the integration variable $\omega_4$ to
	\be
	\omega_4 = \frac{\omega f(z)}{|z-1|} \left| \frac{z_{23}}{z_{24} z_{34}} \right|,
	\ee
	so that Mandelstam invariants are parameterized by functions with zero $\SL(2,\C)$ weight. 	Here we have introduced an arbitrary positive function $f(z)$, which will be chosen later on so as to manifest certain crossing properties.	Collecting all the factors, we find
	\be
	\tilde{\cal A}_4(\epsilon_i, \Delta_i, z_i, \bar{z}_i) = \tfrac{1}{2}\Theta_4(\epsilon_i,z) \prod_{i<j}\left(z_{ij}^{\frac{h}{3} - h_i -h_j} \bar{z}_{ij}^{\frac{\tilde{h}}{3} - \tilde{h}_i - \tilde{h}_j} \right) \tilde{A}_4(\Delta,z),
	\ee
	where all the non-trivial content of the correlator is captured by the function
	\be\label{eq:A4-tilde}
	\boxed{
	\tilde{A}_4(\Delta,z) = \delta(\Im z)\, |z(1{-}z)|^{2-\frac{\Delta}{3}} f(z)^{\Delta-4} \int_{0}^{\infty} \d \omega\, \omega^{\Delta-5} A_4(s^\ast, t^\ast),
	}
	\ee
	which depends only on the sum of conformal dimensions $\Delta= \sum_{i=1}^{n} \Delta_i$ and the real cross-ratio $z$. In terms of the new variables, we have
	\be
	(s^\ast, t^\ast, u^\ast) = 4 \omega^2 f(z)^2 \left( \frac{1}{z{-}1},\; -\frac{1}{z},\; -\frac{1}{z(z{-}1)} \right).
	\ee
	The above expression is independent of the crossing channel and evaluating it in different regions \eqref{eq:4pt-channels} will result in the CCFT correlator in a given crossing channel for any valid choice of $f(z)$. We note that comparing to~\eqref{Acele} above, we are using the $\mathcal{N}_{\Delta_i,\epsilon_i}=1$. Namely, we will see that the object with nice crossing properties at 4-point has no relative phase for the in versus out particles. We will see that this discussion generalizes to $n$-point in what follows.  While this sounds reasonable at the level of the amplitude, it might be somewhat unexpected from the point of view of the bulk wavefunctions, since $\mathcal{N}_{\Delta_i,\epsilon_i}=\Gamma(\Delta_i)^{-1} (i\epsilon_i)^{\Delta_i}$ gives a conformal primary wavefunction where the only difference between incoming and outgoing is the $i\varepsilon$ prescription: $X^0\mapsto X^0 - i \varepsilon \epsilon_i$ for infinitesimal $\varepsilon$.  

	\paragraph{Crossing} Let us now explore how we can exploit the freedom in choosing $f(z)$ to manifest crossing properties of $\tilde{A}_4(\Delta,z)$. Because we expect singularities from the collinear behavior at $z=\{0,1,\infty\}$ we will start with the ansatz $f(z) = |z|^a |1{-}z|^b$ with $a$, $b$ undetermined. Recall that by crossing of the plane-wave amplitude $A(s,t)$, we simply mean switching of all the labels $(\epsilon_i, J_i, \Delta_i, \omega_i, z_i, \bar{z}_i)$ of a particle $i$ with those of $j$. For instance, to go from the $s$-channel to $t$-channel we relabel $1 \leftrightarrow 3$. At the level of the CCFT, this corresponds to $z \to 1-z$. Under this change, we would like to impose
	\be
	(s^\ast, t^\ast, u^\ast) \xrightarrow{z \;\to\; 1-z} (t^\ast, s^\ast, u^\ast). 
	\ee
	This fixes $a=b$. Similarly, to go from the $s$-channel to $u$-channel we use $z \to 1/z$, under which we demand
	\be
	(s^\ast, t^\ast, u^\ast) \xrightarrow{z \;\to\; 1/z} (u^\ast, t^\ast, s^\ast).
	\ee
	This yields $a=b=1/3$. With this choice, we find
	\be\label{eq:A4-crossing}
	\tilde{A}_4(\Delta,z) = \hat{\delta}(\Im z) \int_{0}^{\infty} \d \omega\, \omega^{\Delta-5} A_4\left(\tfrac{4\omega^2 |z(1{-}z)|^{2/3}}{z-1},- \tfrac{4\omega^2 |z(1{-}z)|^{2/3}}{z}\right),
	\ee
	where we have defined the crossing-symmetric delta function
	\be
	\hat{\delta}(\Im z) = |z(1{-}z)|^{2/3}\, \delta(\Im z),
	\ee
	which is invariant under the transformations $z \to 1{-}z$ and $z \to 1/z$.
	
	The expression \eqref{eq:A4-crossing} makes manifest the fact that the CCFT correlator $\tilde{A}_4(\Delta,z)$ inherits crossing properties of the plane-wave amplitude $A_4(s,t)$. For instance, when scattering four identical particles we have
	\be
	A_4(s,t) = A_4(t,s) = A_4(u,t).
	\ee
	The above derivation shows that this is reflected in the CCFT crossing via
	\be
	\tilde{A}_4(\Delta, z) = \tilde{A}_4(\Delta, 1{-}z) = \tilde{A}_4(\Delta, 1/z).
	\ee
	While this property holds independently of the choice of $f(z)$, for the choice used in~\eqref{eq:A4-crossing} it also holds for the $\hat{\delta}$-stripped part, which we will turn to next.

	\paragraph{Analyticity in $\boldsymbol{z}$}
	At this stage one may wonder if the function $\tilde{A}_4(\Delta,z)$, after stripping away the overall $\hat{\delta}(\Im z)$, can be analytically extended from the real line to a holomorphic function of $z \in \CP^1$ on the whole celestial sphere. One motivation for pursuing this question is to exploit complex-analytic properties to derive new constraints on the correlator analogous to dispersion relations in the plane-wave basis. 	Alternatively, one might want to aim for an analytic extension of the whole function $\tilde{\A}_4(\Delta,z)$ to $(z,\tilde{z}) \in \CP^1 \times \CP^1$, where the celestial sphere corresponds to the locus $\tilde{z} = \bar{z}$ and the circle to $z = \bar{z}$. This is the type of continuation needed to understand the connection to $(2,2)$ signature amplitudes (see, e.g., \cite{Atanasov:2021oyu}). It is inherently non-unique, because a given choice of analytic continuation can depend on arbitrarily-complicated functions of the difference $z-\tilde{z}$. 
	
	While various versions of analytic continuations were studied in \cite{Lam:2017ofc,Arkani-Hamed:2020gyp,Chang:2021wvv}, there are essentially two sources of problems that need to be understood before making a general statement beyond tree-level toy models. The first is that upon complexifying $z$, the $\omega$-integral is multi-valued because $A_4(s,t)$ has poles and branch cuts. Actually, this causes problems even at tree-level, because the $\omega$-integrand itself has a branch cut for $\Delta \notin \Z$ (the spurious square-root branching between $\omega$ and the Mandelstam invariants can be removed by a change of variables $\omega \to \sqrt{\omega}$ prior to analytic continuation). The second issue comes from the overall monodromies of the prefactor, such as the ones in \eqref{eq:A4-tilde}. 

\paragraph{Analyticity in $\boldsymbol{\Delta}$} The analyticity in $z$ is simplest in the case of tree-level  scattering in a purely-massless theory. In this case, the stripped amplitudes are rational in the spinor helicity variables, so that the only monodromies can come from the prefactor we have stripped out when defining $\tilde{A}_n$.
However it is also the case that these stripped amplitudes are homogeneous in the overall energy scale.  For example,
	\be
	A_4(\lambda^2 s, \lambda^2 t) = \lambda^{d} A_4(s,t),
	\ee
	for some scale degree $d$ set by the mass dimensions of the external particles. The price we pay for good analyticity in $z$ is a non-analyticity in the remaining invariant quantity $\Delta$, since for such amplitudes
\be
\tilde{A}(\Delta,z)\propto \boldsymbol{\delta}(i(\Delta-4+d))
\ee
where the distribution $\boldsymbol{\delta}$, studied in \cite{Donnay:2020guq}, reduces to a Dirac delta function for real values of the argument. Going beyond tree level we return to the more complicated $z$-dependence, but land on an object that is analytic in $\Delta$ whose pole structure can be matched onto low-energy effective field theory coefficients if one ignores IR problems related to massless exchanges~\cite{Arkani-Hamed:2020gyp,Chang:2021wvv}.

	\subsection{More particles}
		We would now like to extract an explicit form for the invariant quantity $\tilde{A}_n(\epsilon_i,\Delta_i,r_{ijkl})$ in~\eqref{eq:Atilde} for generic $n$, and discuss how to represent the independent invariant data, paying attention to the $\epsilon_i$ dependence.	We will start by generalizing the construction in~\cite{Schreiber:2017jsr} in a manner that makes contact with our celestial circle story.  
		
	For five or more particles, we are able to solve the momentum conservation constraints~\eqref{eq:momem_cons} for four of the energy variables $\omega_I$. In~\eqref{eq:4x4minors} we encountered the $4\times 4$ minors
		\be
	\mathcal{U}_{ijkl}=\det Q_{ijkl},~~~Q_{ijkl}:=\Big(\epsilon_i q_i~ \epsilon_j q_j~ \epsilon_k q_k~ \epsilon_l q_l \Big)
	\ee
which evaluate to
		\be
	\U_{ijkl} = 8\epsilon_i\epsilon_j\epsilon_k\epsilon_l|z_{ik}z_{jl}|^2\, \mathrm{Im} r_{ijkl}.
	\ee
	We also saw that the signs of these minors depend upon the 3-particle celestial circles, as illustrated in Fig.~\ref{4ptcirc}. The momentum conserving delta function constraint then reduces to
	\be\label{eq:omega}
~~	\delta^4({\textstyle\sum}_{i=1}^{n} p_i) = \frac{1}{|\U_{1234}|} \prod_{I=1}^{4} \delta(\omega_I - \omega_I^\ast),~~~~
	\omega_I^\ast = -\frac{1}{\U_{1234}} \sum_{i=5}^{n}  \omega_i \U_{1234/.I\to i},
	\ee
	where we have chosen $I=1,2,3,4$ as a spanning set of reference momenta.  More generally we can pick any set of four particles with linearly independent $q_i$ and use the momentum conserving constraints to localize the corresponding $\omega_i$. Using~\eqref{eq:omega}, the celestial amplitude $\tilde{\A}_n(\epsilon_i, \Delta_i, z_i, \bar{z}_i)$ reduces to
	\be\label{eq:Aloc}\scalemath{.96}{
	\tilde{\A}_n = \frac{1}{|\U_{1234}|} \prod_{i<j} \left( \frac{z_{ij}}{\bar{z}_{ij}} \right)^{\frac{1}{n-2}(\frac{J}{n-1} - J_i - J_j)}
	 \left(\prod_{i=5}^{n} \int_0^\infty  \d \omega_i\, \omega_i^{\Delta_i-1}\right) \prod_{I=1}^{4} (\omega_I^\ast)^{\Delta_i - 1} A_n(s_{ij}^\ast)\Theta(\omega_I^*).
}	\ee
	Note that $\omega_I^\ast$'s are still a function of the $\omega_i$'s with $i>4$. The support of the correlator in the $z_i$-space is not obvious at the level of the integrand precisely because the constraints $\Theta(\omega_i^\ast)$ still have this $\omega_i$ dependence. The integral in~\eqref{eq:Aloc} is over {\it all} such $\omega_{i\ge 5}$ such that $\omega_I^*>0$.    From our discussion of the channel support in Sec.~\ref{sec:kine} we know that the final result is proportional to the constraint $\Theta_n(\epsilon_i,r_{ijk})$, since this defines the region of puncture configurations for which there exists {\it any} such solution.

One drawback of this presentation of the celestial amplitude is that the frequency variables appearing in~\eqref{eq:Aloc} transform non-trivially under $\SL(2,\mathbb{C})$. We can strip off the canonical powers of $z_{ij}$ and $\bz_{ij}$ in~\eqref{eq:Atilde} to extract the Lorentz invariant part $\tilde{A}_n$ by making the following change of variables to a set of $\SL(2,\C)$-invariant `energies' $\Omega_i$
	\be\label{eq:varchange}
		\omega_i =\Omega_i \prod_{\substack{j,k=1\\
	j\neq k\neq i}}^{n} \bigg|\frac{z_{jk}}{z_{ji}z_{ki}}\bigg|^{\frac{1}{(n-1)(n-2)}}.
	\ee
This symmetric prescription avoids making a choice of particular reference punctures by taking a geometric mean of all possible choices of reference punctures.  One can check that 
	\be\label{prodom}
	\prod_{i=1}^{n} \omega_i^{\Delta_i} =
	\Big(\prod_{i=1}^{n} {\Omega}_i^{\Delta_i}\Big) \prod_{i<j} \big|z_{ij}\big|^{\frac{2}{n-2}(\frac{\Delta}{n-1}-\Delta_i-\Delta_j)}.
	\ee
Comparing~\eqref{eq:Atilde} and~\eqref{eq:Aloc} we find
\be\label{atilden}
\tilde{A}_n(\epsilon_i,\Delta_i,r_{ijkl})=\frac{1}{|\U_{1234}|}\prod_{i=5}^n\int_0^\infty  \d \Omega_i\, {\Omega}_i^{\Delta_i-1} \prod_{I=1}^{4} ({\Omega}_I^{\ast})^{\Delta_i - 1} \left(\frac{{\Omega}_I^{\ast}}{\omega_I^\ast}\right)A_n(s_{ij}^\ast)\Theta(\Omega_I^*).
\ee
Here the ratio $\Omega_I^*/\omega_I^*$ only depends on the $|z_{ij}|$ and serves to modify the Jacobian $|\U_{1234}|^{-1}$ that comes from using the momentum conservation delta function to localize the $\omega_I^*$, to the one appropriate for localizing the $\Omega_I^*$. Since it is straightforward to introduce the change of variables~\eqref{eq:varchange} in the $s_{ij}^\ast$, we are set to start analyzing the invariant amplitude $\tilde{A}_n$.

Before doing so, we note that other choices of Lorentz invariant energy variables that differ from $\Omega_i$ by functions of the invariant cross ratios, such as the ones in~\cite{Schreiber:2017jsr}, can also be used to define an invariant amplitude.  These amount to stripping off different kinematical factors.  More explicitly, the choice of $\Omega_i$ in~\eqref{eq:varchange} is appropriate for the canonical form~\eqref{eq:Atilde}, used here and in~\cite{Pasterski:2017ylz,Arkani-Hamed:2020gyp,Brandhuber:2021nez}, while other choices of kinematical factors are common in the CFT literature and can be useful for examining the celestial conformal block decomposition~\cite{Nandan:2019jas,Atanasov:2021cje,Fan:2021isc,Fan:2021pbp,Fan:2022vbz,Hu:2022syq}. 

\paragraph{Invariant Data}
We would now like to identify the free data and crossing channel dependence of $\tilde{A}_n$. Plugging~\eqref{caln} into~\eqref{compact}, we see that besides the $\d\log\omega_i$ measure and the momentum conserving delta function, everything is phrased in terms of the Mandelstam invariants \footnote{ We can also recast this constraint in terms of Lorentz invariants. Imposing momentum conservation amounts to demanding that $(1,1,\ldots,1)$ is in the right null space of the $4\times n$ matrix $p^\mu_j$. This is at most rank four, so if we can form an invertible $4\times 4$ matrix from a choice of $p_{\mu,{I}}$, requiring $(1,1,\ldots,1)$ to be in the kernel of the $4\times n$ sub-matrix $s_{Ij}$ is equivalent. 
 }
\be\label{omint}
\Theta_n\tilde{A}_n
= \int \prod_{i=1}^{n} \frac{\d\omega_i}{\omega_i} \prod_{i<j}\left(-\frac{\epsilon_i\epsilon_j}{4}s_{ij}\right)^{-\frac{2}{n-2}(\frac{\Delta}{n-1} - \Delta_i -\Delta_j)} A_n(s_{ij})\,\delta^{4}({\textstyle\sum}_{i=1}^{n} p_i).
\ee
Recall that the number of independent kinematic invariants in four space-time dimensions is $3n-10$: starting from the $4n$ Lorentz-vector components of the external momenta $p_i^\mu$ we impose $n$ on shell conditions, in addition to $10$ constraints coming from Poincar\'e invariance ($6$ for the choice of Lorentz frame and $4$ from momentum conservation). Alternatively, one can start with the matrix of $n(n-3)/2$ Mandelstam invariants $s_{ij}$ and impose the vanishing of every $5\times 5$ minor, leading to the same number $3n-10$.

This degree of freedom counting should of course carry over to the celestial basis, and we would like to express $\tilde{A}_n$ as a function of $3n-10$ independent variables. Since this stripped celestial amplitude is $\SL(2,\C)$-invariant, it can written as a function of $n-3$ \emph{complex} cross ratios $r_{ijkl}$ for $n\geq 5$. In addition, the conformal dimensions $\Delta_i$ give $n$ continuous variables. Finally, there are $4$ constraints coming from translation invariance. Note that these can be convoluted constraints since translation invariance is no longer manifest on the celestial sphere. In total, this counting leaves us with $3n-10$ degrees of freedom $(r_{ijkl}, \Delta_i)$, which are the counterparts of the independent Mandelstam invariants on the celestial sphere.

Let us now return to the problem of reducing the $\Delta_i$ dependence down to $n-4$ parameters for $n\ge 5$.\footnote{
The independent variables are distributed  slightly differently for  $n=4$ because the external momenta do not span the whole 4D space. As we have seen in Sec.~\ref{sec:four}, the momentum conserving delta function restricted the complex cross ratio $z$ to be real, and provided a constraint on the $\Delta_i$ dependence such that only $\Delta=\sum_{i=1}^4\Delta_i$ appeared as the invariant datum. This gives the expected $3n-10 = 2$ real degrees of freedom.}   Again this is all coming from the fact that translation invariance imposes $4$ constraints on the $\omega_i$ dependence while we are still performing an $n$-dimensional integral transform to the $\Delta_i$. Noting that the translation generators act on our conformal primaries as follows~\cite{Stieberger:2018onx}
\be
P^\mu=\sum_{i=1}^{n} \epsilon_i q_i^\mu e^{\p_{\Delta_i}},
\qquad{\rm where}\qquad e^{\p_{\Delta_i}}:\Delta_i\mapsto\Delta_i+1,
\ee
we can define the action on the invariant amplitude via
\be
P^\mu {\cal A}_n=\Theta_n(\epsilon_i,r_{ijkl}) \prod_{i<j}\left(z_{ij}^{\frac{2}{n-2}(\frac{h}{n-1} - h_i -h_j)} \bar{z}_{ij}^{\frac{2}{n-2}(\frac{\tilde{h}}{n-1} - \tilde{h}_i - \tilde{h}_j)} \right) \tilde{P}^\mu \tilde{A}_n(\epsilon_i,\Delta_i,r_{ijkl}),
\ee
where
\be\badat{3}\label{omderiv}
\tilde{P}^\mu {\tilde A}_n
&=\sum_{k=1}^{n} \epsilon_k q_k^\mu\prod_{\substack{i,j=1\\
	i\neq j\neq k}}^{n} \bigg|\frac{z_{ij}}{z_{jk}z_{ik}}\bigg|^{\frac{1}{(n-1)(n-2)}}e^{\p_{\Delta_k}} \tilde{A}_n.
 \eadat \ee
 The product of $|z_{ij}|$'s comes from commuting the weight shifting operator through the kinematical prefactor. 
From this expression we see that the frequency variables act like weight shifting operators
\be
e^{\p_{\Delta_k}}\tilde{\cal A}_n=\omega_k \tilde{\cal A}_n~~\mapsto~~~ e^{\p_{\Delta_k}}\tilde{A}_n=\Omega_k \tilde{A}_n .
\ee
This is consistent with the fact that the energy variables $\Omega_i$ are Mellin-conjugate to the $\Delta_i$ dependence of the invariant amplitude $\tilde{A}_n$ as in~\eqref{atilden}.

Let us now simplify these constraints, keeping in mind that we would like to write them in a Lorentz invariant form that manifestly reduces the invariant data $\tilde{A}_n$ depends on.
Given this goal, it is easier to start with the Mandelstam invariants
\be
s_{ij}\tilde{\cal A}_n=-4\epsilon_i\epsilon_j |z_{ij}|^2e^{\p_{\Delta_i}+\p_{\Delta_j}}\tilde{\cal A}_n
\ee
whose action on the stripped data $\tilde{A}_n$ we can deduce from step similar to those above
\be\badat{3}\label{sijAn}
s_{ij}\tilde{A}_n&=-4\epsilon_i\epsilon_j|z_{ij}|^2\prod_{\substack{k,l=1\\ k\neq l\neq i}}^{n}\bigg|\frac{z_{kl}}{z_{ki}z_{li}}\bigg|^{\frac{1}{(n-1)(n-2)}}\prod_{\substack{p,q=1\\ p\neq q\neq j}}^{n} \bigg|\frac{z_{pq}}{z_{pj}z_{qj}}\bigg|^{\frac{1}{(n-1)(n-2)}}e^{\p_{\Delta_i}+\p_{\Delta_j}} \tilde{A}_n\\
\eadat\ee
which we can further simplify to an expression involving the cross-ratios:
\be\label{eq:sij-constraints}
s_{ij}\tilde{A}_n =-4\epsilon_i\epsilon_j \bigg(\prod_{\substack{k,l=1\\ k\neq l\neq i}}^{n} \left| r_{ijkl} \right|^{\frac{2}{(n-1)(n-2)}} \bigg) e^{\p_{\Delta_i}+\p_{\Delta_j}} \tilde{A}_n.
\ee
 Holding $i$ fixed and summing over $j$ gives us a differential constraint on $\tilde{A}_n$ as a function of the $\Delta_i$ and cross ratios.  Picking four punctures such that $\U_{i_1i_2i_3i_4}$ is non-vanishing is enough to give us the necessary 4 constraints on the $\Delta_i$ dependence, reducing us to the expected $n-4$ free weights.

\paragraph{Imprints of Crossing Symmetry}

Finally, let us discuss imprints of crossing symmetry of the momentum space amplitude in the celestial basis. As outlined in the introduction, there are two distinct notions of crossing symmetry. In the $\mathcal{S}$-matrix literature it refers to an analytic continuation between crossing channels \cite{Bros:1965kbd}, which has not been demonstrated for massless theories. This question seems more difficult to analyze in the CCFT setup, mainly because our understanding of analyticity of the momentum stripped amplitude $A_n$ as a function of the Mandelstam invariants does not automatically translate to statements about the analyticity of the celestial stripped amplitude as a function of complexified $(\Delta_i,z_i,\bz_i)$. Most likely, a good notion of holomorphic factorization, or conformal-block decomposition, will be needed. For recent progress on CCFT conformal blocks, see, e.g., \cite{Nandan:2019jas,Atanasov:2021cje,Fan:2021isc,Fan:2021pbp,Fan:2022vbz,Hu:2022syq}. 

From the CFT perspective, we might alternatively ask about invariance of correlation functions under exchange of operators. It is rather straightforward to see that CCFT correlators always have this symmetry: if it was present in the momentum space amplitude $A_n(s_{ij})$, it is also in the celestial correlator $\tilde{A}_n(\epsilon_i, \Delta_i, r_{ijkl})$, because the integration kernel in~\eqref{omint} already manifests this symmetry. To be more precise, if the former is invariant under exchanging \emph{all} the labels of the $j$-th and $k$-th particle, then so is the latter. It is an exact statement, not relying on perturbation theory, that is manifest in the extrapolate dictionary~\cite{Pasterski:2021dqe,Donnay:2022sdg}, where it follows from the way the correlators are constructed from the same bulk field pushed to the conformal boundary. A simple example was given in \eqref{eq:fg}. Note that under this symmetry, the support of the celestial correlator $\Theta_n(\epsilon_i)$ will change, because we have exchanged $\epsilon_j \leftrightarrow \epsilon_k$.  For $n \geq 6$, we can consider correlators related by exchange symmetry that have the same support.  However, we stress that in general, even though $\deg(n)$ different correlators can have support at the same point on the celestial sphere, they will be given by distinct functions labelled by $\vec{\epsilon}$. This is consistent with the different OPEs between in-in, out-out, and in-out operators, which leads us to think of the $\epsilon_i$ as an additional label for celestial operators in Lorentizan signature.

\subsection*{Acknowledgements}
\vspace{-1mm}

We would like to thank  Nima Arkani-Hamed, Paolo Benincasa, Freddy Cachazo, Scott Collier, Matthew Heydeman, Atul Sharma, and Herman Verlinde for useful conversations. The research of SM is supported by Frank and Peggy Taplin, as well as the grant DE-SC0009988 from the US Department of Energy.  The research of SP has been supported by the Sam B. Treiman Fellowship at the Princeton Center for Theoretical Science. Research at the Perimeter Institute is supported by the Government of Canada through the Department of Innovation, Science and Industry Canada and by the Province of Ontario through the Ministry of Colleges and Universities.

\appendix

\section{Proving a Useful Theorem}\label{app:thm}

Here we will review the proof of Thm.~2.2 from~\cite{pjm/1103044952}, making  an effort to emphasize here that we can phrase everything in terms of vector spaces and dual vector spaces rather than equipping $\mathbb{R}^m$ with a Euclidean metric. This is useful because in the context of momentum conservation, $n$ corresponds to the number of external particles while $m=4$ is the spacetime dimension.

\setcounter{theorem}{0}
\begin{theorem}[Jackson (2.2) \cite{pjm/1103044952}]\label{thm1}
For an $m\times n$ matrix $Q$, the following are equivalent
\begin{enumerate}[(i)]
    \item $Q{\omega}=0$ has no positive solution ${\omega}\ge0$. 
    \item There exist $v$ such that $v Q>0$.
\end{enumerate}
\end{theorem}

\paragraph{Proof} Let us verify this equivalence in steps. 
 \vspace{1em}
 
\noindent    $\boldsymbol{  \neg \mathrm{(i)} \Rightarrow \neg \mathrm{(ii):}}$~~ This direction is straightforward. Say that there exists a positive ${\omega}\ge 0$ that is not identically 0 such that $Q \omega=0$. Then  for any $v$ in the dual space we have $v Q \omega=0$. Namely there is no $v$ such that $v Q> 0$ since $\omega$ will always be in the null space. 

\vspace{1em}

 \noindent $\boldsymbol{  ~~~ \mathrm{(i)} \Rightarrow ~~ \mathrm{(ii):}}$ ~~ Consider a basis for $\mathbb{R}^n$ defining the positive orthant
    \be
    e_i=(0,\ldots,1,\ldots,0)
    \ee
    with a $1$ in the $i$-th slot.
 Any $\omega\ge 0$ is a positive sum of the $e_i$.  The image of the positive orthant under $Q$ is also a convex set
        \be
    U=\{u| ~\mathrm{for~some}~ \omega\ge0,~u=Q\omega\}
    \ee 
    which is clear by linearity since if $u_1=Q \omega_1$   and $u_2=Q \omega_2$ then
    \be
    \lambda u_1 +(1-\lambda) u_2=Q(\lambda \omega_1 +(1-\lambda)\omega_2).
    \ee
    This set is also closed under positive rescalings and thus is a closed convex cone. The fact that $Q\omega=0$ has no positive solutions means that no non-zero point in the orthant gets mapped to $0$. 

   It also implies that this cone does not contain a linear subspace. If $U$ contained a linear subspace then there would be a point $u\in U$ such that $-u\in U$, or equivalently $u\in -U$ as well.   By linearity, the negative orthant maps to $-U$. If there were a $u$ such that $u\in U$ and $u\in -U$, there would exist an $\omega$ in the positive orthant and an $\varpi$ in the negative orthant such that
   \be 
   Q\omega=Q\varpi \Rightarrow Q(\omega-\varpi)=0
   \ee 
   where $\omega-\varpi>0$. This is disallowed by (i).  Such a convex cone $U\cap -U=\{0\}$ is called {\it salient}.
   
   The next step uses a theorem from Gerstenhaber~\cite{Gerstenhaber}. To prove (ii) we need to show that the dual cone 
    \be
   U^*=\{v|~ vu~\ge 0,~~\forall~ u\in U\}
   \ee
   contains an interior point.  This is guaranteed to be the case if $U^*$ is full-dimensional, which we can show by using the fact that $(U^*)^*=U$ for a closed convex cone.  If $U^*$ were not full dimensional there would be a nonzero $x$ such that 
   \be
   vx=0~~~\forall ~ v\in U^*.
   \ee
   Thus $x\in (U^*)^*$.  However we also have $-x\in  (U^*)^*$.  This cannot be the case if $(U^*)^*=U$ since $U$ is salient. 
   
    Intuitively this is saying that we can find a hyperplane through the origin in $\mathbb{R}^m$ which only intersects this cone at $0$, so that $U$ is contained in a half space.  The dual vector defining this hyperplane will give us the $v$ we need for (ii) since if $vu>0$ for all $u$ in this cone, then $vQ x>0$ for all $x$ in the positive orthant of $\mathbb{R}^n$.  In particular $vQe_i>0$ for all $i$ and so $vQ>0$.

\section{The Lorentz Basis: Then and Now}\label{app:LorentzBasis}

In this appendix we give a guide to the old literature on expressing the ${\cal S}$-matrix in the Lorentz basis, which provides a representation-theoretic perspective complementary to the modern work on celestial amplitudes. It dates back to the work of Joos who considered different bases for one-particle states \cite{Joos:1962qq}, before the plane-wave basis diagonalizing translations became ubiquitous in quantum field theory computations.

Recall that the universal cover of the Poincar\'e group in four dimensions is the semi-direct product $\R^{1,3} \rtimes \mathrm{SL}(2,\C)$ of the group of translations and the Lorentz group. Its (unitary) irreducible representations are one-particle states. The classification of starts with distinguishing between orbits of the first Casimir, $P_\mu P^\mu = m^2 \mathbb{1}$, giving the usual massive ($m^2 > 0$), massless ($m^2 = 0$), tachyon ($m^2 < 0$), and zero-momentum ($P^\mu = 0$) states. Each of these states has additional quantum numbers, which are eigenvalues of the corresponding little group: $\SU(2)$, $\ISO(2) \cong \R^2 \rtimes \mathrm{U}(1)$, $\SU(1,1)$, and $\SL(2,\C)$ respectively, see \cite{Weinberg:1995mt} for the standard reference. The idea is to study particle states as \emph{induced representations} of the Poincar\'e group, but instead of $\R^{1,3}$ (definite-momentum states), using $\SL(2,\C)$ as the basis (definite-boost states). For this purpose, let us first briefly recall the representation theory of $\SL(2,\C)$ itself. See, e.g., \cite{gel2014integral,gelfand2018representations} for classic textbooks.

Calling $L_{\mu\nu}$ the generators of $\SL(2,\C)$, its two Casimirs are given by
\begin{align}
\frac{1}{2} L_{\mu\nu} L^{\mu\nu} &= [J^2 + (\Delta-1)^2 - 1] \mathbb{1},\\
\frac{1}{4!} \epsilon^{\mu\nu\rho\sigma} L_{\mu\nu} L_{\rho\sigma} &= [J(\Delta-1)] \mathbb{1}.
\end{align}
The unitary representations then fall into three classes with the spin: the principal series ($J \in \tfrac{1}{2}\Z$ and $\Delta = 1 + i\lambda$ with $\lambda \in \R$), supplementary series ($J=0$ and $\Delta \in (0,2)\setminus \{1\}$), discrete series ($J=0$ and $\Delta \in \Z\setminus \{1\}$), and the trivial representation ($J=0$ and $\Delta=1$). Only the last two can be finite-dimensional. In order to construct a basis of $\SL(2,\C)$, we need to choose a subgroup, such as $\SU(2)$, $\ISO(2)$, or $\SU(1,1)$. The most common choice are representations induced from the Borel subgroup $B = \ISO(2) \otimes D$ of $\SL(2,\C)$ (i.e., the subgroup of matrices of the form $\left(\begin{smallmatrix}
	a & 0 \\
	c & d 
\end{smallmatrix}\right)$ with $ad=1$), where $D$ is the group of dilations. Following Gel'fand and Neimark's $z$-basis construction \cite{gel1947unitary}, representations in this basis are understood as operators acting on the space of wavefunctions $\psi_{\Delta,J}(z,\bar{z})$ defined on the homogeneous space $\SL(2,\C)/B$. Under the action of the Lorentz group $U[g]$, they transform according to
\be
U[g] \psi_{\Delta,J}(z,\bar{z}) = (cz+d)^{\Delta+J} (\bc \bz+\bd)^{\Delta-J}\psi_{\Delta,J}(\tfrac{az+b}{cz+d},\tfrac{\ba\bar{z}+\bb}{\bc\bar{z}+\bd})
\ee
where $g = \left(\begin{smallmatrix}
	a & b \\
	c & d 
\end{smallmatrix}\right) \in \SL(2,\C)$ with $ad-bc=1$.

Returning back to the Poincar\'e group, the idea is to represent one-particle states as induced representation from $\SL(2,\C)$. Those associated to the $\ISO(2) \otimes D$ subgroup seem particularly convenient for massless states because $\ISO(2)$ is already the little group of massless particles and $D$ leaves the direction of the null momentum unaffected. This means, in the modern language, that points $z$ are identified with coordinates on the celestial sphere. These ideas were applied to the $\mathcal{S}$-matrix by Chakrabarti et al. \cite{doi:10.1063/1.1664709,doi:10.1063/1.1665810} (for $\SU(2)$-induced basis), as well as MacDowell and Roskies \cite{Macdowell:1972ef} (see also \cite{SHAPIRO1962253}), who computed matrix elements and integral transforms to the plane-wave basis. Unfortunately, no explicit scattering amplitudes were studied in the Lorentz basis at that stage. An operator formalism unifying the principal, supplementary, and discrete series was described in \cite{doi:10.1063/1.1665946}. Textbooks on relevant topics include \cite{gel2014integral,gelfand2018representations,ruhl1970lorentz,carmeli2000group,naimark2014linear}.

Of course, $\ISO(2)\otimes D$ is only one choice for constructing a Lorentz basis. It seems natural, for example, to consider $\SU(2)$ when scattering massive particles. For future reference, here we collect references involving explicit results for particular choices of bases:
\be
\begin{tabular}{c|c}
	Basis of $\SL(2,\C)$ & References \\ 
	\hline
	$\ISO(2)\otimes D$ & \cite{Macdowell:1972ef,doi:10.1063/1.1665946,doi:10.1063/1.1664752,doi:10.1063/1.1665299,Novozhilov1969,Bars:1971jil}  \\    
	$\SU(2) \cong \mathrm{SO}(3)$ & \cite{doi:10.1063/1.1664709,doi:10.1063/1.1665810,SHAPIRO1962253,doi:10.1063/1.523705,doi:10.1063/1.1704120,doi:10.1063/1.1704991,doi:10.1142/9789814280389_0009,popov1960theory,doi:10.1142/9789814280389_0019,Paciello1973,Weidemann:1978zv,doi:10.1063/1.1664607,Rühl1969,doi:10.1063/1.524025,PhysRevD.19.3413}  \\  
	$\SU(1,1)$ & \cite{doi:10.1063/1.1665717,doi:10.1063/1.1665764,doi:10.1063/1.1666671}
\end{tabular}
\ee
It would be fascinating to further study the physical significance of these bases in light of the renewed interest in the Lorentz basis.

 \bibliographystyle{utphys}
 \bibliography{references}

\providecommand{\href}[2]{#2}\begingroup\raggedright\begin{thebibliography}{10}

\bibitem{Pasterski:2021raf}
S.~Pasterski, M.~Pate, and A.-M. Raclariu, ``{Celestial Holography},'' in {\em
  {2022 Snowmass Summer Study}}.
\newblock 11, 2021.
\newblock \href{http://arxiv.org/abs/2111.11392}{{\ttfamily arXiv:2111.11392
  [hep-th]}}.

\bibitem{Maldacena:1997re}
J.~M. Maldacena, ``{The Large N limit of superconformal field theories and
  supergravity},'' \href{http://dx.doi.org/10.1023/A:1026654312961}{{\em Adv.
  Theor. Math. Phys.} {\bfseries 2} (1998) 231--252},
  \href{http://arxiv.org/abs/hep-th/9711200}{{\ttfamily arXiv:hep-th/9711200}}.

\bibitem{Witten:1998qj}
E.~Witten, ``{Anti-de Sitter space and holography},''
  \href{http://dx.doi.org/10.4310/ATMP.1998.v2.n2.a2}{{\em Adv. Theor. Math.
  Phys.} {\bfseries 2} (1998) 253--291},
\href{http://arxiv.org/abs/hep-th/9802150}{{\ttfamily arXiv:hep-th/9802150
  [hep-th]}}.

\bibitem{Brown:1986nw}
J.~D. Brown and M.~Henneaux, ``{Central Charges in the Canonical Realization of
  Asymptotic Symmetries: An Example from Three-Dimensional Gravity},''
  \href{http://dx.doi.org/10.1007/BF01211590}{{\em Commun. Math. Phys.}
  {\bfseries 104} (1986) 207--226}.

\bibitem{Brown:1992br}
J.~D. Brown and J.~W. York, Jr., ``{Quasilocal energy and conserved charges
  derived from the gravitational action},''
  \href{http://dx.doi.org/10.1103/PhysRevD.47.1407}{{\em Phys. Rev. D}
  {\bfseries 47} (1993) 1407--1419},
  \href{http://arxiv.org/abs/gr-qc/9209012}{{\ttfamily arXiv:gr-qc/9209012}}.

\bibitem{Bondi:1962px}
H.~Bondi, M.~G.~J. van~der Burg, and A.~W.~K. Metzner, ``{Gravitational waves
  in general relativity. 7. Waves from axisymmetric isolated systems},''
\href{http://dx.doi.org/10.1098/rspa.1962.0161}{{\em Proc. Roy. Soc. Lond.}
  {\bfseries A269} (1962) 21--52}.

\bibitem{Sachs:1962wk}
R.~K. Sachs, ``{Gravitational waves in general relativity. 8. Waves in
  asymptotically flat space-times},''
\href{http://dx.doi.org/10.1098/rspa.1962.0206}{{\em Proc. Roy. Soc. Lond.}
  {\bfseries A270} (1962) 103--126}.

\bibitem{Sachs:1962zza}
R.~Sachs, ``{Asymptotic symmetries in gravitational theory},''
\href{http://dx.doi.org/10.1103/PhysRev.128.2851}{{\em Phys. Rev.} {\bfseries
  128} (1962) 2851--2864}.

\bibitem{Barnich:2009se}
G.~Barnich and C.~Troessaert, ``{Symmetries of asymptotically flat 4
  dimensional spacetimes at null infinity revisited},''
  \href{http://dx.doi.org/10.1103/PhysRevLett.105.111103}{{\em Phys. Rev.
  Lett.} {\bfseries 105} (2010) 111103},
\href{http://arxiv.org/abs/0909.2617}{{\ttfamily arXiv:0909.2617 [gr-qc]}}.

\bibitem{Barnich:2011ct}
G.~Barnich and C.~Troessaert, ``{Supertranslations call for superrotations},''
  {\em PoS} {\bfseries CNCFG} (2010) 010,
  \href{http://arxiv.org/abs/1102.4632}{{\ttfamily arXiv:1102.4632 [gr-qc]}}.
[Ann. U. Craiova Phys.21, S11 (2011)].

\bibitem{Cachazo:2014fwa}
F.~Cachazo and A.~Strominger, ``{Evidence for a New Soft Graviton Theorem},''
\href{http://arxiv.org/abs/1404.4091}{{\ttfamily arXiv:1404.4091 [hep-th]}}.

\bibitem{Kapec:2014opa}
D.~Kapec, V.~Lysov, S.~Pasterski, and A.~Strominger, ``{Semiclassical Virasoro
  symmetry of the quantum gravity $ \mathcal{S}$-matrix},''
  \href{http://dx.doi.org/10.1007/JHEP08(2014)058}{{\em JHEP} {\bfseries 08}
  (2014) 058},
\href{http://arxiv.org/abs/1406.3312}{{\ttfamily arXiv:1406.3312 [hep-th]}}.

\bibitem{Kapec:2016jld}
D.~Kapec, P.~Mitra, A.-M. Raclariu, and A.~Strominger, ``{2D Stress Tensor for
  4D Gravity},'' \href{http://dx.doi.org/10.1103/PhysRevLett.119.121601}{{\em
  Phys. Rev. Lett.} {\bfseries 119} no.~12, (2017) 121601},
\href{http://arxiv.org/abs/1609.00282}{{\ttfamily arXiv:1609.00282 [hep-th]}}.

\bibitem{Eden:1966dnq}
R.~J. Eden, P.~V. Landshoff, D.~I. Olive, and J.~C. Polkinghorne, {\em {The
  analytic S-matrix}}.
\newblock Cambridge Univ. Press, Cambridge, 1966.

\bibitem{Arkani-Hamed:2006emk}
N.~Arkani-Hamed, L.~Motl, A.~Nicolis, and C.~Vafa, ``{The String landscape,
  black holes and gravity as the weakest force},''
  \href{http://dx.doi.org/10.1088/1126-6708/2007/06/060}{{\em JHEP} {\bfseries
  06} (2007) 060}, \href{http://arxiv.org/abs/hep-th/0601001}{{\ttfamily
  arXiv:hep-th/0601001}}.

\bibitem{Arkani-Hamed:2020blm}
N.~Arkani-Hamed, T.-C. Huang, and Y.-T. Huang, ``{The EFT-Hedron},''
  \href{http://dx.doi.org/10.1007/JHEP05(2021)259}{{\em JHEP} {\bfseries 05}
  (2021) 259}, \href{http://arxiv.org/abs/2012.15849}{{\ttfamily
  arXiv:2012.15849 [hep-th]}}.

\bibitem{Arkani-Hamed:2021ajd}
N.~Arkani-Hamed, Y.-t. Huang, J.-Y. Liu, and G.~N. Remmen, ``{Causality,
  Unitarity, and the Weak Gravity Conjecture},''
  \href{http://arxiv.org/abs/2109.13937}{{\ttfamily arXiv:2109.13937
  [hep-th]}}.

\bibitem{Pasterski:2017kqt}
S.~Pasterski and S.-H. Shao, ``{Conformal basis for flat space amplitudes},''
  \href{http://dx.doi.org/10.1103/PhysRevD.96.065022}{{\em Phys. Rev.}
  {\bfseries D96} no.~6, (2017) 065022},
\href{http://arxiv.org/abs/1705.01027}{{\ttfamily arXiv:1705.01027 [hep-th]}}.

\bibitem{ss}
S.~Pasterski, ``{Soft Shadows},'' {\em
  \href{https://physicsgirl.com/ss.pdf}{978-0-9863685-4-7}} (2017) .

\bibitem{Atanasov:2021cje}
A.~Atanasov, W.~Melton, A.-M. Raclariu, and A.~Strominger, ``{Conformal Block
  Expansion in Celestial CFT},''
  \href{http://arxiv.org/abs/2104.13432}{{\ttfamily arXiv:2104.13432
  [hep-th]}}.

\bibitem{Sharma:2021gcz}
A.~Sharma, ``{Ambidextrous light transforms for celestial amplitudes},''
  \href{http://arxiv.org/abs/2107.06250}{{\ttfamily arXiv:2107.06250
  [hep-th]}}.

\bibitem{Ball:2019atb}
A.~Ball, E.~Himwich, S.~A. Narayanan, S.~Pasterski, and A.~Strominger,
  ``{Uplifting AdS$_{3}$/CFT$_{2}$ to flat space holography},''
  \href{http://dx.doi.org/10.1007/JHEP08(2019)168}{{\em JHEP} {\bfseries 08}
  (2019) 168}, \href{http://arxiv.org/abs/1905.09809}{{\ttfamily
  arXiv:1905.09809 [hep-th]}}.

\bibitem{Pasterski:2022lsl}
S.~Pasterski and H.~Verlinde, ``{Chaos in Celestial CFT},''
  \href{http://arxiv.org/abs/2201.01630}{{\ttfamily arXiv:2201.01630
  [hep-th]}}.

\bibitem{deBoer:2003vf}
J.~de~Boer and S.~N. Solodukhin, ``{A Holographic reduction of Minkowski
  space-time},'' \href{http://dx.doi.org/10.1016/S0550-3213(03)00494-2}{{\em
  Nucl. Phys.} {\bfseries B665} (2003) 545--593},
\href{http://arxiv.org/abs/hep-th/0303006}{{\ttfamily arXiv:hep-th/0303006
  [hep-th]}}.

\bibitem{Pasterski:2017ylz}
S.~Pasterski, S.-H. Shao, and A.~Strominger, ``{Gluon Amplitudes as 2d
  Conformal Correlators},''
  \href{http://dx.doi.org/10.1103/PhysRevD.96.085006}{{\em Phys. Rev.}
  {\bfseries D96} no.~8, (2017) 085006},
\href{http://arxiv.org/abs/1706.03917}{{\ttfamily arXiv:1706.03917 [hep-th]}}.

\bibitem{Pate:2019lpp}
M.~Pate, A.-M. Raclariu, A.~Strominger, and E.~Y. Yuan, ``{Celestial Operator
  Products of Gluons and Gravitons},''
  \href{http://arxiv.org/abs/1910.07424}{{\ttfamily arXiv:1910.07424
  [hep-th]}}.

\bibitem{Guevara:2021abz}
A.~Guevara, E.~Himwich, M.~Pate, and A.~Strominger, ``{Holographic Symmetry
  Algebras for Gauge Theory and Gravity},''
  \href{http://arxiv.org/abs/2103.03961}{{\ttfamily arXiv:2103.03961
  [hep-th]}}.

\bibitem{Strominger:2021lvk}
A.~Strominger, ``{w(1+infinity) and the Celestial Sphere},''
  \href{http://arxiv.org/abs/2105.14346}{{\ttfamily arXiv:2105.14346
  [hep-th]}}.

\bibitem{Himwich:2021dau}
E.~Himwich, M.~Pate, and K.~Singh, ``{Celestial Operator Product Expansions and
  ${\rm w}_{1+\infty}$ Symmetry for All Spins},''
  \href{http://arxiv.org/abs/2108.07763}{{\ttfamily arXiv:2108.07763
  [hep-th]}}.

\bibitem{Atanasov:2021oyu}
A.~Atanasov, A.~Ball, W.~Melton, A.-M. Raclariu, and A.~Strominger, ``{(2, 2)
  Scattering and the celestial torus},''
  \href{http://dx.doi.org/10.1007/JHEP07(2021)083}{{\em JHEP} {\bfseries 07}
  (2021) 083}, \href{http://arxiv.org/abs/2101.09591}{{\ttfamily
  arXiv:2101.09591 [hep-th]}}.

\bibitem{Crawley:2021auj}
E.~Crawley, A.~Guevara, N.~Miller, and A.~Strominger, ``{Black Holes in Klein
  Space},'' \href{http://arxiv.org/abs/2112.03954}{{\ttfamily arXiv:2112.03954
  [hep-th]}}.

\bibitem{Mizera:2021ujs}
S.~Mizera, ``{Bounds on Crossing Symmetry},''
  \href{http://dx.doi.org/10.1103/PhysRevD.103.L081701}{{\em Phys. Rev. D}
  {\bfseries 103} no.~8, (2021) 081701},
  \href{http://arxiv.org/abs/2101.08266}{{\ttfamily arXiv:2101.08266
  [hep-th]}}.

\bibitem{Mizera:2021fap}
S.~Mizera, ``{Crossing symmetry in the planar limit},''
  \href{http://dx.doi.org/10.1103/PhysRevD.104.045003}{{\em Phys. Rev. D}
  {\bfseries 104} no.~4, (2021) 045003},
  \href{http://arxiv.org/abs/2104.12776}{{\ttfamily arXiv:2104.12776
  [hep-th]}}.

\bibitem{HM}
H.~S. Hannesdottir and S.~Mizera, ``{What is the $i\varepsilon$ for the
  S-matrix?},'' {\em {to appear}} .

\bibitem{Law:2019glh}
Y.~T.~A. Law and M.~Zlotnikov, ``{Poincar\'e Constraints on Celestial
  Amplitudes},''
\href{http://arxiv.org/abs/1910.04356}{{\ttfamily arXiv:1910.04356 [hep-th]}}.

\bibitem{Arkani-Hamed:2020gyp}
N.~Arkani-Hamed, M.~Pate, A.-M. Raclariu, and A.~Strominger, ``{Celestial
  Amplitudes from UV to IR},''
  \href{http://arxiv.org/abs/2012.04208}{{\ttfamily arXiv:2012.04208
  [hep-th]}}.

\bibitem{Pasterski:2021dqe}
S.~Pasterski, A.~Puhm, and E.~Trevisani, ``{Revisiting the Conformally Soft
  Sector with Celestial Diamonds},''
  \href{http://arxiv.org/abs/2105.09792}{{\ttfamily arXiv:2105.09792
  [hep-th]}}.

\bibitem{Donnay:2022sdg}
L.~Donnay, S.~Pasterski, and A.~Puhm, ``{Goldilocks Modes and the Three
  Scattering Bases},'' \href{http://arxiv.org/abs/2202.11127}{{\ttfamily
  arXiv:2202.11127 [hep-th]}}.

\bibitem{Schreiber:2017jsr}
A.~Schreiber, A.~Volovich, and M.~Zlotnikov, ``{Tree-level gluon amplitudes on
  the celestial sphere},''
  \href{http://dx.doi.org/10.1016/j.physletb.2018.04.010}{{\em Phys. Lett.}
  {\bfseries B781} (2018) 349--357},
\href{http://arxiv.org/abs/1711.08435}{{\ttfamily arXiv:1711.08435 [hep-th]}}.

\bibitem{pjm/1103044952}
{Jackson, James}, ``On the existence problem of linear programming,'' {\em
  {Pacific Journal of Mathematics}} {\bfseries {4}} no.~{1}, ({1954}) {29 --
  36}.

\bibitem{Pasterski:2022jzc}
S.~Pasterski, ``{A Shorter Path to Celestial Currents},''
  \href{http://arxiv.org/abs/2201.06805}{{\ttfamily arXiv:2201.06805
  [hep-th]}}.

\bibitem{OEIS}
N.~J.~A. Sloane, ``{Sequence A000125/M1100 },'' {\em
  \href{https://oeis.org/A000125}{The On-Line Encyclopedia of Integer
  Sequences}} .

\bibitem{Brandhuber:2021nez}
A.~Brandhuber, G.~R. Brown, J.~Gowdy, B.~Spence, and G.~Travaglini,
  ``{Celestial Superamplitudes},''
  \href{http://arxiv.org/abs/2105.10263}{{\ttfamily arXiv:2105.10263
  [hep-th]}}.

\bibitem{Chang:2021wvv}
C.-M. Chang, Y.-t. Huang, Z.-X. Huang, and W.~Li, ``{Bulk locality from the
  celestial amplitude},'' \href{http://arxiv.org/abs/2106.11948}{{\ttfamily
  arXiv:2106.11948 [hep-th]}}.

\bibitem{Lam:2017ofc}
H.~T. Lam and S.-H. Shao, ``{Conformal Basis, Optical Theorem, and the Bulk
  Point Singularity},''
\href{http://arxiv.org/abs/1711.06138}{{\ttfamily arXiv:1711.06138 [hep-th]}}.

\bibitem{Donnay:2020guq}
L.~Donnay, S.~Pasterski, and A.~Puhm, ``{Asymptotic Symmetries and Celestial
  CFT},'' \href{http://dx.doi.org/10.1007/JHEP09(2020)176}{{\em JHEP}
  {\bfseries 09} (2020) 176}, \href{http://arxiv.org/abs/2005.08990}{{\ttfamily
  arXiv:2005.08990 [hep-th]}}.

\bibitem{Nandan:2019jas}
D.~Nandan, A.~Schreiber, A.~Volovich, and M.~Zlotnikov, ``{Celestial
  Amplitudes: Conformal Partial Waves and Soft Limits},''
  \href{http://dx.doi.org/10.1007/JHEP10(2019)018}{{\em JHEP} {\bfseries 10}
  (2019) 018},
\href{http://arxiv.org/abs/1904.10940}{{\ttfamily arXiv:1904.10940 [hep-th]}}.

\bibitem{Fan:2021isc}
W.~Fan, A.~Fotopoulos, S.~Stieberger, T.~R. Taylor, and B.~Zhu, ``{Conformal
  Blocks from Celestial Gluon Amplitudes},''
  \href{http://arxiv.org/abs/2103.04420}{{\ttfamily arXiv:2103.04420
  [hep-th]}}.

\bibitem{Fan:2021pbp}
W.~Fan, A.~Fotopoulos, S.~Stieberger, T.~R. Taylor, and B.~Zhu, ``{Conformal
  Blocks from Celestial Gluon Amplitudes II: Single-valued Correlators},''
  \href{http://arxiv.org/abs/2108.10337}{{\ttfamily arXiv:2108.10337
  [hep-th]}}.

\bibitem{Fan:2022vbz}
W.~Fan, A.~Fotopoulos, S.~Stieberger, T.~R. Taylor, and B.~Zhu, ``{Elements of
  Celestial Conformal Field Theory},''
  \href{http://arxiv.org/abs/2202.08288}{{\ttfamily arXiv:2202.08288
  [hep-th]}}.

\bibitem{Hu:2022syq}
Y.~Hu, L.~Lippstreu, M.~Spradlin, A.~Y. Srikant, and A.~Volovich, ``{Four-point
  correlators of light-ray operators in CCFT},''
  \href{http://arxiv.org/abs/2203.04255}{{\ttfamily arXiv:2203.04255
  [hep-th]}}.

\bibitem{Stieberger:2018onx}
S.~Stieberger and T.~R. Taylor, ``{Symmetries of Celestial Amplitudes},''
  \href{http://dx.doi.org/10.1016/j.physletb.2019.03.063}{{\em Phys. Lett. B}
  {\bfseries 793} (2019) 141--143},
  \href{http://arxiv.org/abs/1812.01080}{{\ttfamily arXiv:1812.01080
  [hep-th]}}.

\bibitem{Bros:1965kbd}
J.~Bros, H.~Epstein, and V.~Glaser, ``{A proof of the crossing property for
  two-particle amplitudes in general quantum field theory},''
  \href{http://dx.doi.org/10.1007/BF01646307}{{\em Commun. Math. Phys.}
  {\bfseries 1} no.~3, (1965) 240--264}.

\bibitem{Gerstenhaber}
{Gerstenhaber, Murray}, ``Theory of convex polyhedral cones,'' {\em {Chap.
  XVIII of Cowles Commission Monograph Activity analysis of production and
  allocation, {\rm ed. T.C. Koopmans}}} no.~{13}, ({1951}) {298 -- 316}.

\bibitem{Joos:1962qq}
H.~Joos, ``{On the Representation theory of inhomogeneous Lorentz groups as the
  foundation of quantum mechanical kinematics},''
  \href{http://dx.doi.org/10.1002/prop.2180100302}{{\em Fortsch. Phys.}
  {\bfseries 10} (1962) 65--146}.

\bibitem{Weinberg:1995mt}
S.~Weinberg, {\em {The Quantum theory of fields. Vol. 1: Foundations}}.
\newblock Cambridge University Press, 6, 2005.

\bibitem{gel2014integral}
I.~Gel'fand, M.~Graev, and N.~Vilenkin, {\em Generalized Functions, Volume 5:
  Integral Geometry and Representation Theory}.
\newblock Elsevier Science, 2014.

\bibitem{gelfand2018representations}
I.~Gel'fand, R.~Minlos, and Z.~Shapiro, {\em Representations of the Rotation
  and Lorentz Groups and Their Applications}.
\newblock Dover Publications, 2018.

\bibitem{gel1947unitary}
I.~M. Gel'fand and M.~A. Naimark, ``{Unitary representations of the Lorentz
  group},'' {\em Izvestiya Rossiiskoi Akademii Nauk. Seriya Matematicheskaya}
  {\bfseries 11} no.~5, (1947) 411--504.

\bibitem{doi:10.1063/1.1664709}
A.~Chakrabarti, M.~Levy‐Nahas, and R.~Seneor, ``{`Lorentz Basis' of the
  Poincaré Group},'' \href{http://dx.doi.org/10.1063/1.1664709}{{\em Journal
  of Mathematical Physics} {\bfseries 9} no.~8, (1968) 1274--1283}.

\bibitem{doi:10.1063/1.1665810}
A.~Chakrabarti, ``{Lorentz Basis of the Poincaré Group II},''
  \href{http://dx.doi.org/10.1063/1.1665810}{{\em Journal of Mathematical
  Physics} {\bfseries 12} no.~9, (1971) 1822--1840}.

\bibitem{Macdowell:1972ef}
W.~W. Macdowell and R.~Roskies, ``{Reduction of the Poincare group with respect
  to the Lorentz group},'' \href{http://dx.doi.org/10.1063/1.1665882}{{\em J.
  Math. Phys.} {\bfseries 13} (1972) 1585--1591}.

\bibitem{SHAPIRO1962253}
I.~Shapiro, ``Expansion of the scattering amplitude in relativistic spherical
  functions,'' \href{http://dx.doi.org/10.1016/0031-9163(62)91370-7}{{\em
  Physics Letters} {\bfseries 1} no.~7, (1962) 253--255}.

\bibitem{doi:10.1063/1.1665946}
I.~Bars and F.~Gursey, ``{Operator Treatment of the Gel'fand‐Naimark Basis
  for SL(2,C)},'' \href{http://dx.doi.org/10.1063/1.1665946}{{\em Journal of
  Mathematical Physics} {\bfseries 13} no.~2, (1972) 131--143}.

\bibitem{ruhl1970lorentz}
W.~Ruhl, {\em {The Lorentz Group and Harmonic Analysis}}.
\newblock Mathematical physics monograph series. W. A. Benjamin, 1970.

\bibitem{carmeli2000group}
M.~Carmeli, {\em {Group Theory and General Relativity: Representations of the
  Lorentz Group and Their Applications to the Gravitational Field}}.
\newblock World Scientific, 2000.

\bibitem{naimark2014linear}
M.~Naimark and H.~Farahat, {\em {Linear Representations of the Lorentz Group}}.
\newblock ISSN. Elsevier Science, 2014.

\bibitem{doi:10.1063/1.1664752}
S.~Chang and L.~O'Raifeartaigh, ``{Unitary Representations of SL(2, C) in an
  E(2) Basis},'' \href{http://dx.doi.org/10.1063/1.1664752}{{\em Journal of
  Mathematical Physics} {\bfseries 10} no.~1, (1969) 21--29}.

\bibitem{doi:10.1063/1.1665299}
G.~J. Iverson and G.~Mack, ``{$\overline{E2}$-Parametrization of SL(2, C)},''
  \href{http://dx.doi.org/10.1063/1.1665299}{{\em Journal of Mathematical
  Physics} {\bfseries 11} no.~5, (1970) 1581--1584}.

\bibitem{Novozhilov1969}
Y.~V. Novozhilov and E.~V. Prokhvatilov, ``{Representations of the Poincar{\'e}
  group in E(2) bases},'' \href{http://dx.doi.org/10.1007/BF01028573}{{\em
  Theoretical and Mathematical Physics} {\bfseries 1} no.~1, (Oct, 1969)
  78--93}.

\bibitem{Bars:1971jil}
I.~Bars and F.~Guersey, ``{Duality and the Lorentz group},''
  \href{http://dx.doi.org/10.1103/PhysRevD.4.1769}{{\em Phys. Rev. D}
  {\bfseries 4} (1971) 1769--1776}.

\bibitem{doi:10.1063/1.523705}
G.~B. Smith, ``Matrix element expansion of a spin wavefunction,''
  \href{http://dx.doi.org/10.1063/1.523705}{{\em Journal of Mathematical
  Physics} {\bfseries 19} no.~3, (1978) 581--585}.

\bibitem{doi:10.1063/1.1704120}
J.~S. Lomont and H.~E. Moses, ``{The Representations of the Inhomogeneous
  Lorentz Group in Terms of an Angular Momentum Basis},''
  \href{http://dx.doi.org/10.1063/1.1704120}{{\em Journal of Mathematical
  Physics} {\bfseries 5} no.~2, (1964) 294--298}.

\bibitem{doi:10.1063/1.1704991}
J.~S. Zmuidzinas, ``{Unitary Representations of the Lorentz Group on 4‐Vector
  Manifolds},'' \href{http://dx.doi.org/10.1063/1.1704991}{{\em Journal of
  Mathematical Physics} {\bfseries 7} no.~4, (1966) 764--780}.

\bibitem{doi:10.1142/9789814280389_0009}
K.-C. Chou and L.~G. Zastavenko,
  \href{http://dx.doi.org/10.1142/9789814280389_0009}{``{The Shapiro Integral
  Transformation},''} in {\em Selected Papers of K C Chou}, pp.~33--38.

\bibitem{popov1960theory}
V.~Popov, ``On the theory of the relativistic transformations of the wave
  functions and density matrix of particles with spin,'' {\em Soviet Physics
  JETP} {\bfseries 37} no.~10, (1960) .

\bibitem{doi:10.1142/9789814280389_0019}
K.-C. Chou and L.~G. Zastavenko,
  \href{http://dx.doi.org/10.1142/9789814280389_0019}{``{Integral
  Transformations of the I. S. Shapiro Type for Particles of Zero Mass},''} in
  {\em Selected Papers of K C Chou}, pp.~77--80.

\bibitem{Paciello1973}
M.~L. Paciello, A.~Sciarrino, and B.~Taglienti, ``Projective invariance of
  dual-resonance models from spin analyticity and lorentz invariance,''
  \href{http://dx.doi.org/10.1007/BF02756276}{{\em Il Nuovo Cimento A
  (1965-1970)} {\bfseries 14} no.~3, (Apr, 1973) 591--604}.

\bibitem{Weidemann:1978zv}
A.~W. Weidemann, ``{Quantum Fields in a `Lorentz Basis'},''
  \href{http://dx.doi.org/10.1007/BF02776455}{{\em Nuovo Cim. A} {\bfseries 57}
  (1980) 221}.

\bibitem{doi:10.1063/1.1664607}
N.~Mukunda, ``{Zero‐Mass Representations of the Poincar{\'e} Group in an O(3,
  1) Basis},'' \href{http://dx.doi.org/10.1063/1.1664607}{{\em Journal of
  Mathematical Physics} {\bfseries 9} no.~4, (1968) 532--536}.

\bibitem{Rühl1969}
W.~R{\"u}hl, ``{The convolution of Fourier transforms and its application to
  the decomposition of the momentum operator on the homogeneous Lorentz
  group},'' \href{http://dx.doi.org/10.1007/BF02754927}{{\em Il Nuovo Cimento A
  (1971-1996)} {\bfseries 63} no.~4, (Oct, 1969) 1131--1162}.

\bibitem{doi:10.1063/1.524025}
M.~Daumens and M.~Perroud, ``{Internal Lorentz basis for two‐particle
  states},'' \href{http://dx.doi.org/10.1063/1.524025}{{\em Journal of
  Mathematical Physics} {\bfseries 20} no.~12, (1979) 2621--2627}.

\bibitem{PhysRevD.19.3413}
M.~Daumens, M.~Perroud, and P.~Winternitz, ``{Relativistic energy-dependent
  partial-wave analysis for particles with spin},''
  \href{http://dx.doi.org/10.1103/PhysRevD.19.3413}{{\em Phys. Rev. D}
  {\bfseries 19} (Jun, 1979) 3413--3425}.

\bibitem{doi:10.1063/1.1665717}
A.~D. Steiger, ``{Poincaré‐Irreducible Tensor Operators for Positive‐Mass
  One‐Particle States I},'' \href{http://dx.doi.org/10.1063/1.1665717}{{\em
  Journal of Mathematical Physics} {\bfseries 12} no.~7, (1971) 1178--1191}.

\bibitem{doi:10.1063/1.1665764}
A.~D. Steiger, ``{Poincaré‐Irreducible Tensor Operators for Positive‐Mass
  One‐Particle States II},'' \href{http://dx.doi.org/10.1063/1.1665764}{{\em
  Journal of Mathematical Physics} {\bfseries 12} no.~8, (1971) 1497--1507}.

\bibitem{doi:10.1063/1.1666671}
B.~Radhakrishnan and N.~Mukunda, ``{Spacelike representations of the
  inhomogeneous Lorentz group in a Lorentz basis},''
  \href{http://dx.doi.org/10.1063/1.1666671}{{\em Journal of Mathematical
  Physics} {\bfseries 15} no.~4, (1974) 477--490}.

\end{thebibliography}\endgroup

\end{document}